\documentclass[12pt,preprint]{aastex}

\usepackage[english]{babel}
\usepackage[utf8x]{inputenc}
\usepackage[T1]{fontenc}

\usepackage{amsmath,bm}
\usepackage{amssymb}
\usepackage{graphicx}
\usepackage[colorlinks=true, allcolors=blue]{hyperref}
\usepackage{booktabs}
\usepackage{ulem}
\usepackage{color}
\usepackage{amsmath}
\usepackage[symbol]{footmisc}
\usepackage{natbib}
\usepackage{lineno}
\usepackage{epstopdf}

\slugcomment{}

\shorttitle{Secular resonances \& terrestrial planets in binary stars}
\shortauthors{Haghighipour \& Andrew}

\begin{document}

\title{Secular Resonances in Planet-Hosting Binary Stars. I. \\
General Theory}

\author{Nader Haghighipour \altaffilmark{1,2,3} and Michael Andrew  \altaffilmark{4}}
\altaffiltext{1}{Planetary Science Institute, Tucson, AZ, USA} 
\altaffiltext{2}{Institute for Astronomy, University of Hawaii-Manoa,Honolulu, HI, USA}
\altaffiltext{3}{Institute for Advanced Planetary Astrophysics, Honolulu, HI, USA}
\altaffiltext{4}{University of Hawaii-Manoa, Honolulu, HI, USA}
\email{nader@psi.edu}

\begin{abstract}
Motivated by the diversity of circumstellar planets in binary stars and the strong effects of the secular 
resonances of Jupiter and Saturn on the formation and architecture of the inner solar system, we have launched 
an expansive project on studying the effects of secular resonances on the formation of terrestrial planets 
around a star of a moderately close binary. As the first phase of our project, we present here the general 
theory of secular resonances in dual-star systems where the primary hosts two giant planets. Using the 
concept of generalized disturbing function, we derive the formula for the locations of secular resonances 
and show that in systems where the perturbation of the secondary star is stronger, the locations of secular 
resonances are farther way from the primary and closer to the giant planets. The latter implies that in such 
systems, terrestrial planet formation has a larger area to proceed with more of the protoplanetary disk being 
available to it. To demonstrate the validity of our theoretical results, we simulated the evolution of a 
protoplanetary disk interior to the inner giant planet. Results, in addition to confirming our theoretical 
predictions, pointed to an important finding: In binary stars, the perturbation of the secondary suppresses 
the secular resonances of giant planets. Simulations also show that as the disk loses material, secular 
resonances move inward, scattering objects out of the disk and/or facilitating their collisional growth. 
We present results of our study and discuss their implications for the simulations of terrestrial planet formation.
\end{abstract}

\section{Introduction}
The secular perturbation of giant planets have played fundamental roles in the formation and dynamical 
evolution of the inner solar system. As shown by the classic works of \citet{Milani92,Milani94}, 
secular resonances, in particular those of Jupiter and Saturn, have been crucial to shaping the architecture of 
the asteroid belt. Also, as demonstrated by \citet{Levison03} and \citet{Haghighipour16}, these resonances have 
had significant contributions to the final mass, water content, and orbital architecture of terrestrial planets 
including the containment of terrestrial planet formation to the region interior to 2.1 au. 

The detection of circumstellar planets in moderately close binary stars (i.e., binaries with separations/periastrons smaller 
than 40 au) has extended the interest in understanding the effects of secular resonances to a broader context.
Thanks to the successful operations of space and ground-based telescopes, in the past 15 years, 40 of such binaries 
have been identified.\footnote{A complete and up-to-date list of binary stars with planets 
can be found at http://exoplanet.eu/planets\_binary.} In most of these systems,
the binary hosts only one planet. However, in a few cases, two or more planets have been detected around one of
the stars. 

Although majority of these planets have masses in the range of  Neptune-to-Jupiter mass, their mere existence 
suggests that planets in moderately close binaries are diverse and these systems may also host terrestrial-class planets. 
The latter raises the following question: given the effects of the secular resonances of giant planets on the formation and
orbital architecture of the inner solar system, how would these resonances manifest themselves and affect the formation,
composition, and orbital assembly of terrestrial planets when the planet-hosting star is subject to the perturbation of a 
nearby stellar companion? The goal of this two-article series is to address the above question. 
 
We begin by presenting the theory of secular perturbation for planetary systems in moderately close binaries. 
Although our approach is general and can be applied to any binary system with multiple circumstellar
planets, in order to compare the results with those in our solar system, we consider a binary with two giant planets around one of 
the stars. We will develop the theory for the full four-body system and identify the locations and characteristics of the
secular resonances of the two giant planets. In the next paper, we will simulate the formation of terrestrial planets in these 
systems and use the theory presented here to demonstrate the effects of secular resonances on the formation and the composition
of the final bodies. 

The structure of this paper is as follows. In Section 2, we present the complete analytical development of our theory. 
In Section 3, we apply our theory to some planet-hosting binary stars and calculate the locations of their secular resonances.
Section 4 demonstrates the reliability of our theory by comparing its results with those of numerical simulations, and also shows
the effects of the perturbation of a second star on the appearance and evolution of secular resonances. 
Section 5 concludes this portion of the study by presenting a summary of the results and discussing their implications.

\section{Analytical Calculations}
We consider a coplanar system consisting of a moderately close binary with two giant planets orbiting one of the stars.
This assumption implies that planet formation has proceeded constructively 
around this star meaning that, after the star's protoplanetary disk was truncated by the perturbation 
of its stellar companion, the disk maintained enough mass to form planets.
It also implies that the two planets are safe from the perturbation of the secondary star and maintain stable orbits
\citep[see][for comprehensive reviews of the stability of planetary orbits in binary star systems]{Quarles20,Quarles24}. 

Our goal is to determine the locations of the secular resonances of the two giant planets. 
Secular resonances appear when the rate of the variations of the longitude of the pericenter of a 
small body (e.g., a planetesimal or a planetary embryo) becomes commensurate with that of a planet. Because we are
ultimately interested in studying the effects of these resonances on the formation of terrestrial planets, we adopt
a basic approach where the Lagrange's equations of motion are used at the lowest order of approximation and 
the disturbing functions are expanded to the second order in eccentricity. 
A detailed study of secular resonances, including the effect of orbital inclination, requires a complex analysis 
of the system, which is beyond the scope of this study.

\subsection{Secular perturbations of the planets and secondary star}
We begin by considering the four-body system of the binary and two planets.
For the mere sake of identification, we consider the planet-hosting star to be the primary (P) and refer to 
the other stellar component as the secondary (S). We  assume, without any loss of generality, 
that the system is coplanar and place the origin of the coordinates system on the primary star.
In this scenario, the secondary star rotates around the primary in a Keplerian orbit. Figure 1 shows the system. 

Because we are interested in calculating the variations of the longitude of the pericenter, it is more convenient
to express the dynamics of each body using Lagrange's equations of motion. Denoting the inner and outer planets
and the secondary star with indices I, O, and S, respectively, to the lowest order in eccentricity in our coplanar 
system, these equations can be written as

\begin{equation}
{{d{e_j}}\over {dt}} = - {1\over {{n_j}{e_j}{a_j^2}}}{{\partial{R_j}}\over {\partial{\varpi_j}}}
\qquad , \qquad
{{d{\varpi_j}}\over {dt}}= {1\over {{n_j}{e_j}{a_j^2}}}{{\partial{R_j}}\over {\partial{e_j}}}\,.
\end{equation}

\noindent
In these equations, $m_j$, $a_j$, $e_j$ and $\varpi_j$ are the mass, semimajor axis, eccentricity, and 
the longitude of pericenter of the $j$th body, 
${n_j^2}=G({m_{\rm p}}+{m_j}){a_j^{-3}}$ is the body's mean-motion, $m_{\rm p}$ is the mass of the primary star, and $G$ is 
the gravitational constant. The function $R_j$ in equations (1) is the disturbing function and is given by

\begin{equation}
{R_j} = \bigg[{{\mu_k}\over{|{{\bm r}_k}-{{\bm r}_j}|}} + {{\mu_\ell}\over{|{{\bm r}_\ell}-{{\bm r}_j}|}}\bigg]
- \bigg[{\mu_k}{{{{\bm r}_k} \cdot {{\bm r}_j}}\over{r_k^3}} + {\mu_\ell}{{{{\bm r}_\ell} \cdot 
{{\bm r}_j}}\over{r_\ell^3}}\bigg]
\quad , \quad j\neq k\neq \ell= {\rm I,O,S}.
\end{equation}

\noindent
Here ${\mu_j}= G {m_j}$ and ${\bm r}_j$ is the position vector of the $j$th body with respect to the primary star
with ${r_j} = |{\bm r}_j|$. The first bracket in equation (2) shows the direct part of the disturbing function and 
the second brackets, showing the scalar products between two position vectors, is the indirect part. 

It is common to show the solutions to equations (1) using the eccentricity vectors

\begin{equation}
{e_j} \sin {\varpi_j} = {\sum_k} {e_{jk}} \sin ({g_k}t + {\theta_k})
\qquad, \qquad
{e_j} \cos {\varpi_j} = {\sum_k} {e_{jk}} \cos ({g_k}t + {\theta_k})\,,
\end{equation}

\noindent
where $g_k$ represents the rate of the variations of the longitude of pericenter. To calculate this quantity,
we turn our attention to the parts of the disturbing function that govern the secular evolution of the system. 
We note that when the two planets are not in a mean-motion resonance, the secular perturbations 
manifest themselves through those terms of the disturbing function that are independent of mean longitudes. 
These terms appear only in the direct part of the disturbing function as all terms in the indirect part 
contain at least one mean longitude. To the second order in eccentricity and first order in planet/primary 
mass ratio, these terms are given by \citep{Ellis2000}

\begin{align}
& {b_{1/2}^{(0)}}({\alpha_{jk}}),\\
& {1\over 8}\, ({e_j^2}+{e_k^2})\,
\Big[2{\alpha_{jk}}{d\over{d{\alpha_{jk}}}} + {\alpha_{jk}^2} {{d^2}\over{d{\alpha_{jk}^2}}} \Big]
{b_{1/2}^{(0)}({\alpha_{jk}})},\\
& {1\over 4}\, {e_j}{e_k}\,
\Big[2 - 2{\alpha_{jk}}{d\over{d{\alpha_{jk}}}} - {\alpha_{jk}^2} {{d^2}\over{d{\alpha_{jk}^2}}} \Big] 
{b_{1/2}^{(1)}({\alpha_{jk}})}.
\end{align}

\noindent
In these expressions, $\alpha$ is the ratio of the semimajor axis of an inner body to that 
of an outer object, and

\begin{equation}
{b_q^{(p)}}({\alpha_{jk}}) = {1\over \pi}\, {\int_0^{2\pi}}{\cos(p\psi)}
(1-2{\alpha_{jk}}\cos\psi+ {\alpha_{jk}^2})^{-q} d\psi \,,
\end{equation}

\noindent
is a Laplace coefficient. The quantities $p$ and $q$ in this equation are given by $p=1,2,3 ...,$ and 
$q={1/2, 3/2, 5/2, ...}$, respectively. 
Using equation (7), ${b_{1/2}^{(0)}}({\alpha_{jk}})$ can be written as

\begin{equation}
\begin{aligned}[b]
{b_{1/2}^{(0)}}({\alpha_{jk}}) & = 
{(1+{\alpha_{jk}})^{-1}}\, {_2F_1}\big[0.5, 0.5, 1,{{4{\alpha_{jk}}}{(1+{\alpha_{jk}})^{-2}}}\Big]\\
& + {(1-{\alpha_{jk}})^{-1}}\, {_2F_1}\Big[0.5, 0.5, 1,{{-4{\alpha_{jk}}}{(1-{\alpha_{jk}})^{-2}}}\Big],
\end{aligned}
\end{equation}

\noindent
where ${_2F_1}(\beta,\gamma,\delta,z)$ is the hypergeometric function with $\beta, \gamma$ and $\delta$ 
being real numbers. As shown by equation (8), ${b_{1/2}^{(0)}}({\alpha_{jk}})$ is only a function 
of semimajor axis which means, for a given semimajor axis-ratio $(\alpha)$, this quantity will converge into a 
real number and, therefore, will not contribute to the secular evolution of the system. For that reason, we will
no longer include this term in our calculations. Simplifying the terms inside the brackets in equations (5) and (6)
using the identities

\begin{equation}
2\alpha{{d{b_{1/2}^{(0)}}}\over{d\alpha}} + {\alpha^2}{{{d^2}{b_{1/2}^{(0)}}}\over{d{\alpha^2}}}=
\alpha{b_{3/2}^{(1)}}\, ,
\end{equation}

\noindent
and

\begin{equation}
2{b_{1/2}^{(1)}} - 2\alpha{{d{b_{1/2}^{(1)}}}\over{d\alpha}} - {\alpha^2}{{{d^2}{b_{1/2}^{(1)}}}\over{d{\alpha^2}}}=
-\alpha{b_{3/2}^{(2)}}\, ,
\end{equation}

\noindent 
the secular terms of the disturbing functions can be written as

\begin{equation}
\begin{aligned}[b]
{R_{\rm I}}&={n_{\rm I}^2}{a_{\rm I}^2}{\Big({{m_{\rm O}}\over{{m_{\rm P}}+{m_{\rm I}}}}\Big)}
\bigg[{1\over 8}{\alpha_{\rm IO}^2}{e_{\rm I}^2}{b_{3/2}^{(1)}}({\alpha_{\rm IO}})-
{1\over 4}{\alpha_{\rm IO}^2}{{e_{\rm I}}{e_{\rm O}}}{b_{3/2}^{(2)}}({\alpha_{\rm IO}})
\cos({\varpi_{\rm I}}-{\varpi_{\rm O}})\bigg]\\
&+{n_{\rm I}^2}{a_{\rm I}^2}{\Big({{m_{\rm S}}\over{{m_{\rm P}}+{m_{\rm I}}}}\Big)}
\bigg[{1\over 8}{\alpha_{\rm IS}^2}{e_{\rm I}^2}{b_{3/2}^{(1)}}({\alpha_{\rm IS}})-
{1\over 4}{\alpha_{\rm IS}^2}{{e_{\rm I}}{e_{\rm S}}}{b_{3/2}^{(2)}}({\alpha_{\rm IS}})
\cos({\varpi_{\rm I}}-{\varpi_{\rm S}})\bigg],
\end{aligned}
\end{equation}

\begin{equation}
\begin{aligned}[b]
{R_{\rm O}}&={n_{\rm O}^2}{a_{\rm O}^2}{\Big({{m_{\rm I}}\over{{m_{\rm P}}+{m_{\rm O}}}}\Big)}
\bigg[{1\over 8}{\alpha_{\rm IO}}{e_{\rm O}^2}{b_{3/2}^{(1)}}({\alpha_{\rm IO}})-
{1\over 4}{\alpha_{\rm IO}}{{e_{\rm I}}{e_{\rm O}}}{b_{3/2}^{(2)}}({\alpha_{\rm IO}})
\cos({\varpi_{\rm I}}-{\varpi_{\rm O}})\bigg]\\
&+{n_{\rm O}^2}{a_{\rm O}^2}{\Big({{m_{\rm S}}\over{{m_{\rm P}}+{m_{\rm O}}}}\Big)}
\bigg[{1\over 8}{\alpha_{\rm OS}^2}{e_{\rm O}^2}{b_{3/2}^{(1)}}({\alpha_{\rm OS}})-
{1\over 4}{\alpha_{\rm OS}^2}{{e_{\rm O}}{e_{\rm S}}}{b_{3/2}^{(2)}}({\alpha_{\rm OS}})
\cos({\varpi_{\rm O}}-{\varpi_{\rm S}})\bigg],
\end{aligned}
\end{equation}

\begin{equation}
\begin{aligned}[b]
{R_{\rm S}}&={n_{\rm S}^2}{a_{\rm S}^2}{\Big({{m_{\rm I}}\over{{m_{\rm P}}+{m_{\rm S}}}}\Big)}
\bigg[{1\over 8}{\alpha_{\rm IS}}{e_{\rm S}^2}{b_{3/2}^{(1)}}({\alpha_{\rm IS}})-
{1\over 4}{\alpha_{\rm IS}}{{e_{\rm I}}{e_{\rm S}}}{b_{3/2}^{(2)}}({\alpha_{\rm IS}})
\cos({\varpi_{\rm I}}-{\varpi_{\rm S}})\bigg]\\
&+{n_{\rm S}^2}{a_{\rm S}^2}{\Big({{m_{\rm O}}\over{{m_{\rm P}}+{m_{\rm S}}}}\Big)}
\bigg[{1\over 8}{\alpha_{\rm OS}}{e_{\rm S}^2}{b_{3/2}^{(1)}}({\alpha_{\rm OS}})-
{1\over 4}{\alpha_{\rm OS}}{{e_{\rm O}}{e_{\rm S}}}{b_{3/2}^{(2)}}({\alpha_{\rm OS}})
\cos({\varpi_{\rm O}}-{\varpi_{\rm S}})\bigg].
\end{aligned}
\end{equation}

\noindent
Replacing ${e_j^2}$ in the first term inside each bracket with ${e_j} {e_j} \cos({\varpi_j} - {\varpi_j})$,
the above expressions can be written in the following compact form 

\begin{equation}
{R_j} = {1\over 2} {n_j} {a_j^2} \big[{A_{jk}}\big] {e_j} {e_k} \cos({\varpi_j} - {\varpi_k})
\qquad , \qquad j\neq k\neq \ell= {\rm I,O,S}.
\end{equation}

\noindent
In equation (14), $\big[{A_{jk}}\big]$ is a $3 \times 3$ matrix with the components,

\begin{align}
{A_{11}} &= {{n_{\rm I}}\over {4({{m_{\rm p}}+{m_{\rm I}}})}}
\Big[{m_{\rm O}}{\alpha_{\rm {IO}}^2}{b_{3/2}^{(1)}}({\alpha_{\rm {IO}}}) +
{m_{\rm S}}{\alpha_{\rm {IS}}^2}{b_{3/2}^{(1)}}({\alpha_{\rm {IS}}})\Big],\\
{A_{12}} &= -\,{{n_{\rm I}}\over {4({{m_{\rm p}}+{m_{\rm I}}})}}
\Big[{m_{\rm O}}{\alpha_{\rm {IO}}^2}{b_{3/2}^{(2)}}({\alpha_{\rm {IO}}})\Big],\\
{A_{13}} &= -\,{{n_{\rm I}}\over {4({{m_{\rm p}}+{m_{\rm I}}})}}
\Big[{m_{\rm S}}{\alpha_{\rm {IS}}^2}{b_{3/2}^{(2)}}({\alpha_{\rm {IS}}})\Big],\\
{A_{21}} &= -\,{{n_{\rm O}}\over {4({{m_{\rm p}}+{m_{\rm O}}})}}
\Big[{m_{\rm I}}{\alpha_{\rm {IO}}}{b_{3/2}^{(2)}}({\alpha_{\rm {IO}}})\Big],\\
{A_{22}} &= {{n_{\rm O}}\over {4({{m_{\rm p}}+{m_{\rm O}}})}}
\Big[{m_{\rm I}}{\alpha_{\rm {IO}}}{b_{3/2}^{(1)}}({\alpha_{\rm {IO}}}) +
{m_{\rm S}}{\alpha_{\rm {OS}}^2}{b_{3/2}^{(1)}}({\alpha_{\rm {OS}}})\Big],\\
{A_{23}} &= -\,{{n_{\rm O}}\over {4({{m_{\rm p}}+{m_{\rm O}}})}}
\Big[{m_{\rm S}}{\alpha_{\rm {OS}}^2}{b_{3/2}^{(2)}}({\alpha_{\rm {OS}}})\Big],\\
{A_{31}} &= -\,{{n_{\rm S}}\over {4({{m_{\rm p}}+{m_{\rm S}}})}}
\Big[{m_{\rm I}}{\alpha_{\rm {IS}}}{b_{3/2}^{(2)}}({\alpha_{\rm {IS}}})\Big],\\
{A_{32}} &= -\,{{n_{\rm S}}\over {4({{m_{\rm p}}+{m_{\rm S}}})}}
\Big[{m_{\rm O}}{\alpha_{\rm {OS}}}{b_{3/2}^{(2)}}({\alpha_{\rm {OS}}})\Big],\\
{A_{33}} &= {{n_{\rm S}}\over {4({{m_{\rm p}}+{m_{\rm S}}})}}
\Big[{m_{\rm I}}{\alpha_{\rm {IS}}}{b_{3/2}^{(1)}}({\alpha_{\rm {IS}}}) +
{m_{\rm O}}{\alpha_{\rm {OS}}}{b_{3/2}^{(1)}}({\alpha_{\rm {OS}}})\Big],
\end{align}

\noindent
The quantities ${g_k}$ in equations (3) are the eigenvalues of the matrix ${[A_{jk}]}$.

\subsection{Location of secular resonances}
As mentioned earlier, secular resonances occur when the rate of the variations of the longitude of the 
pericenter of a small body becomes commensurate with that of a planet. The locations of these
resonances correspond to the semimajor axes where the longitude of the pericenter of the small body 
becomes equal to $g_k$. 

To determine these locations, we first calculate the rate of the variation of the longitude of 
pericenter for the small body. As in the previous section, we start with the Lagrange's equations of motion.
Denoting the small body by the index (E), these equations are

\begin{equation}
{{d{e_{\rm E}}}\over {dt}} = - {1\over {{n_{\rm E}}{e_{\rm E}}{a_{\rm E}^2}}}{{\partial{R_{\rm E}}}\over {\partial{\varpi_{\rm E}}}},
\qquad , \qquad
{{d{\varpi_{\rm E}}}\over {dt}}= {1\over {{n_{\rm E}}{e_{\rm E}}{a_{\rm E}^2}}}{{\partial{R_{\rm E}}}\over {\partial{e_{\rm E}}}},
\end{equation}

\noindent
where

\begin{equation}
{R_{\rm E}}={\sum_j}{1\over 4}{n_{\rm E}^2}{a_{\rm E}^2}{\Big({{m_j}\over{{m_{\rm P}}+{m_{\rm E}}}}\Big)}
\bigg[{1\over 2}{\alpha_{{\rm E}j}^2}{e_{\rm E}^2}{b_{3/2}^{(1)}}({\alpha_{{\rm E}j}})-
{\alpha_{{\rm E}j}^2}{{e_{\rm E}}{e_j}}{b_{3/2}^{(2)}}({\alpha_{{\rm E}j}})
\cos({\varpi_{\rm E}}-{\varpi_j})\bigg]
\end{equation}

\noindent
is the body's disturbing function due to all three objects exterior to it. The solutions to 
equations (24) are given by 

\begin{equation}
{e_{\rm E}} \sin {\varpi_{\rm E}} = {e_{\rm E}} \sin ({\Omega_{\rm E}}t + {\theta_{\rm E}})\quad, \quad
{e_{\rm E}} \cos {\varpi_{\rm E}} = {e_{\rm E}} \cos ({\Omega_{\rm E}}t + {\theta_{\rm E}})\,,
\end{equation}

\noindent
where

\begin{equation}
{\Omega_{\rm E}}={\sum_j}{1\over 4}{n_{\rm E}}{\Big({{m_j}\over{m_{\rm P}}}\Big)}
{\alpha_{{\rm E}j}^2}{b_{3/2}^{(1)}}({\alpha_{{\rm E}j}})\,,
\end{equation}

\noindent
is the frequency of the variations of the longitude of the pericenter of the small body.
Figure 2 shows ${\Omega}_{\rm E}$ in terms of semimajor axis for a 20 au binary with 
a G-star primary and different secondary stars (see the next section for the choices of the secondary). 
As expected, and shown by $\alpha_{{\rm E}j}$ in equation (27), ${{\Omega}_{\rm E}}$ becomes singular at the semimajor 
axes of the outer perturbers. 
From equations (3) and (27), secular resonances appear where ${{\Omega}_{\rm E}} = {g_k}$.

\section{Application of the theory}

In this section, we use our general theory to determine the locations of secular resonances in
a number of binary stars with circumstellar planets. To demonstrate the validity of our theory and reliability 
of its results, we will present in the next section a comparison between these theoretical predictions and
the outcome of numerical integrations.

Recall that our system consists of a binary with two giant planets orbiting the primary star. For the mere sake of 
simplicity, and in order to be able to compare the results with the effects of the secular resonances of 
Jupiter and Saturn, we assume that the inner planet is Jupiter mass and the outer planet has the mass of Saturn. 
We place the inner planet at the semimajor axis of 1.6 au and the outer planet at 2.94 au. These semimajor axes 
have been chosen to ensure that for all combinations of binary parameters considered in our study, the two planets 
will be inside the binary's stable zone and, therefore, safe from the perturbation of the secondary star
\citep{Quarles20,Quarles24}.
They also correspond to a near 5:2 commensurability, similar to the period ratio of Jupiter and Saturn.
Other orbital elements of the planets are taken to be equal to those of Jupiter and Saturn, respectively.

In order to examine the effect of the secondary star, we consider this star to be of 
M, K, G, and F types with a mass equal to 0.4, 0.7, 1 and 1.3 solar masses, respectively. We vary the
semimajor axis of the binary from 20 to 50 au, and its eccentricity between 0 and 0.5.
Because we are also interested in the formation of terrestrial planets in these systems, and because 
models of terrestrial planet formation have been developed primarily for the formation of these bodies around
the Sun, we consider the primary to be a solar-type G star. 

Tables 1-4 show the elements of the matrix ${[A_{jk}]}$ and their corresponding eigenvalues $(g_k)$
for all combinations of the mass of the secondary star and the semimajor axis of the binary. Calculations were 
carried out in all generality where, despite the fact that due to its large mass, the effect of the secondary star would be 
negligibly small, this star was also included. An inspection of the results points to an interesting finding:
the magnitude of $g_k$ increases with increasing the mass 
of the secondary star and decreasing the binary's semimajor axis. In other words, $g_k$ is larger
in systems where the perturbation of the secondary is stronger. This trend can also be seen in the values of $g_k$
that correspond to the secondary star. Despite their extremely small values, which confirm that no secular resonances
can appear due to the sole effect of the secondary, the magnitudes of these quantities still increase by 
increasing the secondary's effect. 

The increase in the magnitude of $g_k$ has two immediate implications. Because secular resonances appear where ${g_k}={{\Omega}_{\rm E}}$, 
a larger $g_k$ means that 
the initial location of its corresponding secular resonance will appear at a larger semimajor axis (i.e., closer to the inner planet 
and farther away from the primary star).
Figures 3 and 4 show this for a variety of different cases. In these figures, the locations of secular resonances 
have been marked at the semimajor axes where the graphs of the frequencies $g_k$ and ${\Omega}_{\rm E}$ intersect. Figure 3 
demonstrates the effect of binary separation by showing the locations of these
resonances for different values of the semimajor axis in a GG binary, and figure 4 shows the effect of the mass 
of the secondary star by considering a binary with a separation of 40 au. As shown by these figures, the initial locations 
of secular resonances are at larger semimajor axes in binaries with smaller separations and appear farther away 
from the primary star as the mass of the secondary increases.

\section{Comparison with numerical simulations}

\subsection{Confirming the results of the general theory}

To confirm the validity of our theoretical results and to examine the effect of the secondary star,
we simulated the evolution of a protoplanetary disk interior to
the orbit of the inner planet. We considered a disk extending from 0.5 to 1.5 au, consisting of approximately 100 Moon- 
to Mars-sized planetary embryos and over 400 small planetesimals. The surface density of the disk was chosen
to follow an $r^{-1.5}$ profile and planetary embryos were placed randomly at 5--10 mutual Hill radii.
We integrated the entire system including the disk, giant planets, and the secondary star
for 10 Myr using a special purpose integrator developed specifically for orbital integrations 
in binary star systems \citep{Chambers03}.

Figures 5 and 6 show samples of the results for a GM, GK, GG, and GF binary with a separation of 20 au. 
The top panels in these figures show the initial locations of the secular resonances, $({a_{g_1}},{a_{g_2}})$
as obtained from our general theory.  
The bottom panels show the snapshots of the evolution of the disk. The objects in red are planetary embryos and 
those in gray are planetesimals. The two giant planets, not shown in the figure, are in their orbits at 1.6 and 
2.94 au. The dashed lines correspond to the locations of mean-motion resonances with the inner giant planet.
An inspection of these results clearly demonstrates that, in strong agreement with the theoretical
predictions, secular resonances appear either precisely on their predicted locations  
or in their slight vicinity. To guide the reader's eyes, we have marked the locations of $g_1$ and $g_2$ 
by black arrows on the bottom panels. 

We would like to note that, as demonstrated by equations (15) -- (23) and equation (27), the quantities $g_k$ and $\Omega_{\rm E}$
are independent of the eccentricity. Because secular resonance occur where ${g_k} = {\Omega_{\rm E}}$, the locations
of these resonances are also independent of the eccentricity. The latter indicates that the above results and their 
confirmations of the predictions of our general theory are valid for and can be extended to binaries with eccentric 
planets and secondary stars as well.

\subsection{The effect of the secondary star}

\subsubsection{No secular resonances due to the secondary star}

As mentioned before, our theory has been developed in complete generality. In addition to the secular resonances due to 
the giant planets, we have calculated the possibility of the appearance of secular resonances due to the secondary star
as well. However, as shown by tables 1-4, the eigenvalues $g_k$ corresponding to these resonances are extremely small 
indicating that if no giant planet existed, the sole perturbation of the secondary star would not result in secular resonances. 
Figure 7 shows this for a 30 au GG binary with no giant planets. The panel on the left corresponds to a circular binary 
and the one on the right represent a binary with an eccentricity 0.2. For the sake of simplicity, we only show the planetary
embryos. As shown here, during the same integration time 
that the secular resonances of the giant planets manifest themselves in figures 5 and 6, the planetary embryos in these
system maintain their low eccentricities confirming the nonoccurrence of a secular resonance. 
The small increase in the eccentricities of a few of these bodies are due to 
their mutual interactions and the perturbation of the secondary star. As a reminder, the planetary embryos have masses 
ranging from the mass of the Moon to the mass of Mars. The mutual interactions of these bodies can in fact affect their orbits.

\subsubsection{Suppression of secular resonances}

Although no secular resonances appear due to the sole effect of the secondary, this star plays a significant 
role on the strength and manifestation of the secular resonances of the giant planets. The perturbation of 
this body confines the variations of the longitudes of the ascending nodes of the planets into a smaller region
causing the orbital precessions of the embryos that are captured in secular resonances with them to have smaller 
amplitudes as well. The latter results in weaker excitation in the orbits of these embryos manifesting itself as a smaller 
increase in their orbital eccentricities. Figure 8 shows this for the location of $g_2$ for the GK binary of figure 5 and
the GF binary of figure 6. The panels on the top show the first appearance of this resonance in the full four-body system. The
panels below them show the same in a system without the secondary star\footnote{The simulations of the two panels corresponding 
to no secondary star were carried out using the $N$-body integration package, Mercury \citep{Chambers99}. The small differences
between the results of these two panels are because the two simulations were carried out on two different computers.}. 
As shown here, in all cases, the orbits of the
embryos at the location of $g_2$ carry higher eccentricities in the bottom panels strongly indicating 
that in a planet-hosting binary star, the perturbation of the secondary suppresses secular resonance of
the giant planets.

\subsubsection{Protoplanetary disk rapid mass loss}

In general, in a protoplanetary disk with an external giant planet, the perturbation of the planet increases the 
orbital eccentricities of planetesimals and planetary embryos in its neighborhood. This rise in eccentricity intensifies 
the mutual interactions among the disk's bodies causing the increase in their eccentricities to propagate to all objects 
in the disk. The latter has several important consequences. First, it amplifies the rate of the collisions of the disk's
objects with the giant planet and central star. It also causes many of these bodies to be scattered into larger
orbits where they no longer reside within the boundaries of the disk (normally, these objects are considered to have 
been ejected from the system). As a result, the disk loses considerable amount of mass.
Figure 9 shows this for the systems of figures 5 and 6. The green curves in this figure show the evolution of the total
mass of the disk. The yellow, blue, and cyan curves show the rates of the mass loss due to the collision with the primary star, 
the inner planet, and the outer planet, respectively. The red curve shows the loss of mass due to scattering out of the system.
A comparison between this figure and the results shown in \citet[][see their Figure 8]{Haghighipour16} demonstrates 
that in a binary system, the perturbation of the secondary star causes the mass loss to start much earlier and continue 
at much higher rate. As a result, in a planet-hosting binary, the protoplanetary disk loses mass more rapidly than
in a single star system. As explained below, the latter will have direct consequences on the dynamics of secular resonances
and as will be demonstrated in Paper II, it will directly affect the formation of smaller planets in the disk.

\subsubsection{Migration of secular resonances}

The rise in the eccentricity among the disk's bodies enhances the rate of their mutual collision as well. While some of
these collisions result in breakage and fragmentation, which are then lost through the above-mentioned processes,
some result in the growth, producing large planetary embryos. As a result, in a short time, the disk loses significant
amount of mass and contains fewer number of bodies. At this state, small embryos interior to the large ones will be 
subject to the secular perturbation of the latter bodies and are temporarily captured in secular resonances with the larger
embryos. This process too increases the eccentricities of the smaller embryos causing the disk to lose more mass.
The loss of mass by the disk weakens its regressing effect on the bodies,
which, combined with the slight inward migration of the giant planets due to their interactions with the disk, causes the
location of the planet's secular resonance to move inward. The rate of this inward migration, the time of its
appearance, and its duration depend strongly on the initial mass and surface density of the disk. We refer the reader
to \citet{Haghighipour16} for an in-depth explanation of the migration of secular resonances and the demonstration of
the process for disks with different surface density profile and for different initial orbital configurations of Jupiter
and Saturn. 

Similar inward migration of secular resonances occurred in all our systems. The right panel of figure 6 shows 
an example of that for a GF binary with a separation of 20 au. As shown here, in 0.2 Myr, the location of 
$({g_2})$ moved inward by approximately 0.15 au and in 0.3 Myr, $({g_1})$ moved in by 0.08 au until the effects
of both resonances were no longer distinguished from the effect of the mutual interactions of embryos. We will
demonstrate in Paper II that such migration of secular resonances will have direct effects on the formation as
well as the final masses and orbits of the terrestrial planets.
We remind the reader that, as mentioned above, the perturbation of 
the secondary star suppresses secular resonances to the extent that in many systems it will be hard to separate its 
effect in exciting the eccentricities of planetary embryos from that caused by mean-motion resonances and the embryos 
mutual interactions.

\section{Summary and concluding remarks}

As the first phase of an initiative on examining the effects of secular resonances on the 
formation and composition of terrestrial planets in moderately close binary stars, we have developed 
the general theory of secular perturbation for planet-hosting binaries. Applying the theory to a
double-star system consisting of two giant planets around the primary, we have presented the procedure 
for calculating the locations of secular resonances in coplanar systems.
To demonstrate the validity of our theory, we integrated the orbits of a large number of planetesimals and protoplanetary
bodies interior to the orbit of the inner giant planet and showed that, in agreement with our
theory, secular resonances of the two planets appear at or in close vicinity to their predicted values. 

To demonstrate the effect of the secondary star, we calculated the locations of secular resonances 
for different values of the mass of the secondary as well as the binary's semimajor axis and eccentricity. Results pointed
to a very interesting finding: Locations of secular resonances move farther away from the primary star and appear
earlier in binaries where the perturbation of the secondary is stronger. The latter is due to the fact that
in planet-hosting binaries, the giant planets act as a medium for transferring the perturbation of the secondary 
star to the objects interior to them. As the secondary star interacts with a planet, the eccentricity of the
planet increases and so does its close approaches to the objects interior to its orbit enhancing their mutual
interactions. Stronger secondary-planet interaction results in stronger and earlier increase in the planet's eccentricity
which in turn results in stronger interactions between the planet and the embryos. The latter causes the 
embryos closer to the planet to be the first ones to show its effects (normally in the form of increase in their orbital 
eccentricities) including the signs of synchronization between their longitude of periastrons and that of the planet.

As mentioned earlier, we considered our systems to be coplanar. This consideration was made to reduce the level of 
complexities associated with systems where the planetary orbits are inclined with respect to the plane of the binary. 
In these systems, inclination decouples from the eccentricity during the system's secular evolution 
\citep[e.g.,][]{Li14} changing the way that the perturbation of the giant planets affects the dynamics of the protoplanetary 
disk and the outcome of the collisional growth. We will study this effect in Paper II where we will carry out
integrations with inclined planetary orbits and demonstrate the consequences of the results on the final assembly and composition 
of terrestrial planets.

The most important outcome of our study is that in planet-hosting binaries, the perturbation of the secondary star 
suppresses the effects of the secular resonances of the giant planets. This can also be seen in our
solar system where unlike the secular resonance of Saturn, which strongly affects the orbits of asteroids in the 
inner part of the asteroid belt, the secular resonance of Jupiter (influenced by the perturbation of Saturn) only slightly 
perturbs the orbits of mall bodies in the vicinity of 1.9 au. When a second star is added to the system, it is, therefore, 
not surprising that, given the large mass of the secondary, the perturbing effect of the secular resonance of the outer 
giant planet too is suppressed in a short time. 

Results of our integrations also demonstrated that as expected, during the evolution of the system, secular resonances migrate toward
the primary star causing many of the disk's objects to be scattered out of the system. Given that while migrating, the secondary star
suppresses the effects of these resonances, it would be important to determine whether and to what extent these resonances perturb 
the formation and composition of smaller planets. We present the results of such investigations in Paper II.

Finally, we would like to note that 
because the purpose of our numerical integrations was to demonstrate the validity of our general theory, we limited
our analysis to only one value of the semimajor axis of the inner and outer planets. However, given the generality of our
analytical approach, similar results stand if the two planets are in farther or closer orbits. 
At larger distances \citep[barring orbital stability, see e.g.,][]{Quarles20,Quarles24} 
the interaction between the secondary star and planets intensifies causing
the secular resonances to appear earlier and be stronger. As these resonances will also be farther out, it will 
take them longer to migrate and be suppressed. That means, for planets in larger orbits, secular resonances will have more time and
space to disturb the protoplanetary disk. In contrast, if the two planets are in smaller semimajor axes, their interactions
with the secondary star will be weaker and so will be the intensity of their secular resonances. The fact that these resonances
have smaller intensity and appear in orbits closer to the primary star implies that they are suppressed more rapidly
and will not have significant effects on the evolution of the protoplanetary disk. We will study these effects in Paper II 
and show how the 
variations in the intensity of secular resonances due to different values of the semimajor axis of the giant planets affect the 
formation and composition of smaller bodies interior to their orbits.

N.H. acknowledges support through NASA grants 80NSSC18K0519, 80NSSC21K1050, and 80NSSC23K0270 and NSF grant AST-2109285.
We are deeply grateful to the Information Technology division of the Institute for Astronomy at the
University of Hawaii-Manoa for maintaining computational resources that were used for carrying out the numerical
simulations of this study. We would like to thank the anonymous referee for their critically reading our manuscript and 
for their useful suggestions and recommendations.

\clearpage

\begin{figure}
\vskip -60pt
\centerline{\includegraphics[scale=0.6]{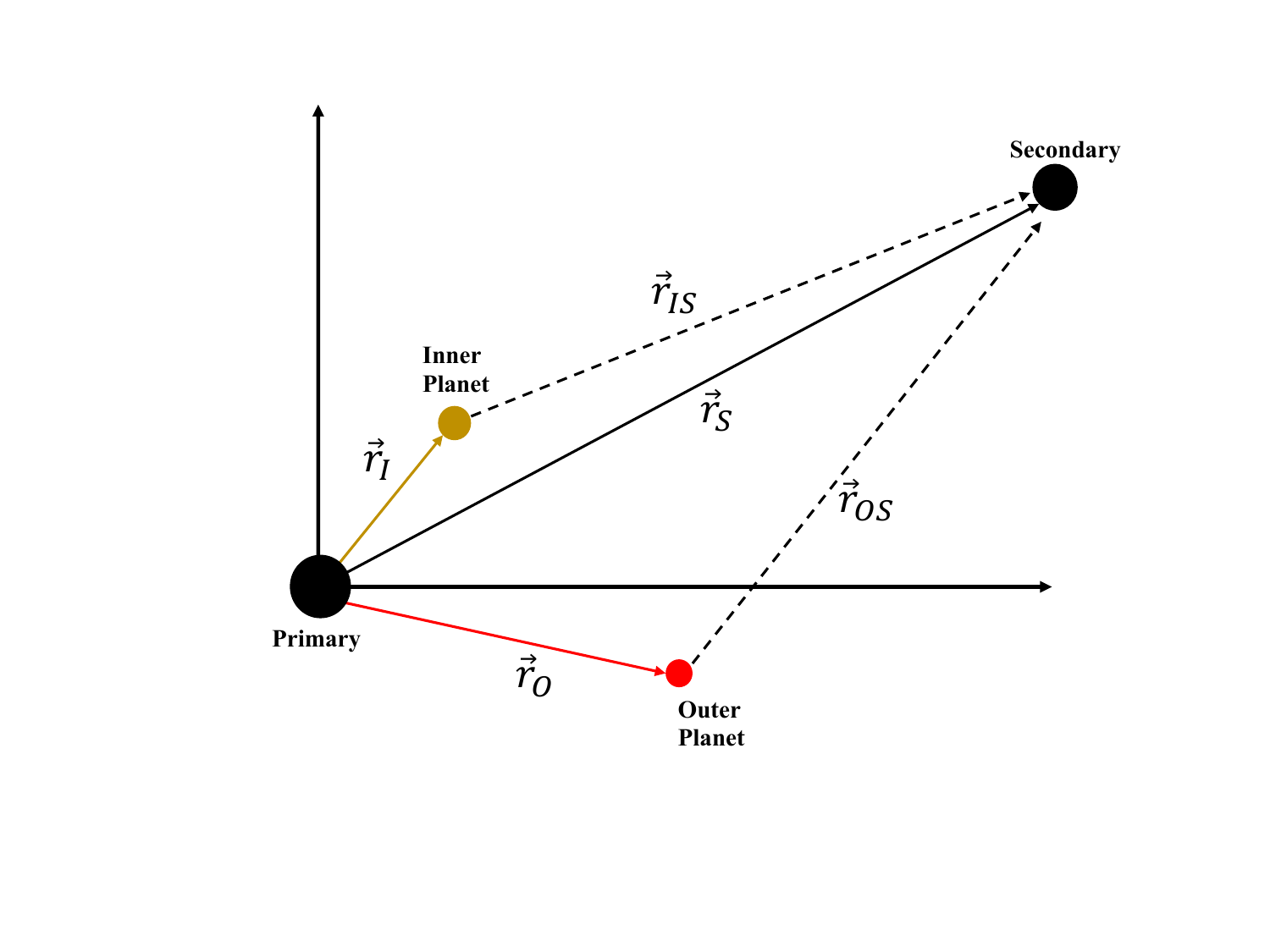}}
\vskip -70pt
\caption{Schematic presentation of the system showing the binary star and the two planets orbiting the primary.
The primary star is at the origin of the coordinate system. The indices I, O, S refer to the inner and
outer planet, and the secondary star, respectively.}
\end{figure}

\clearpage

\begin{figure*}[ht]
\includegraphics[scale=0.9]{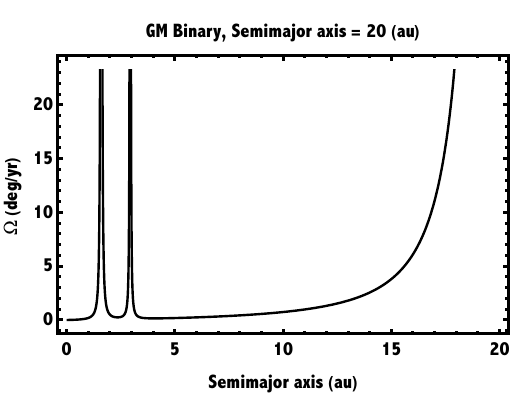}
\includegraphics[scale=0.9]{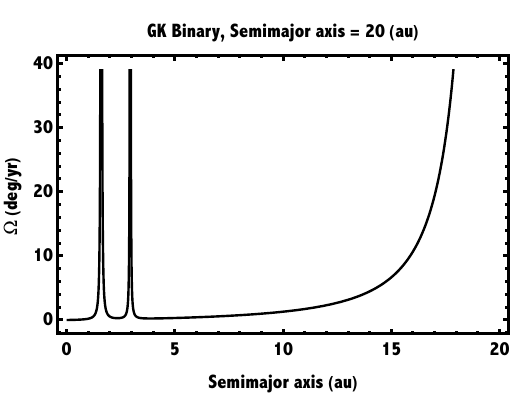}
\vskip 20pt
\includegraphics[scale=0.9]{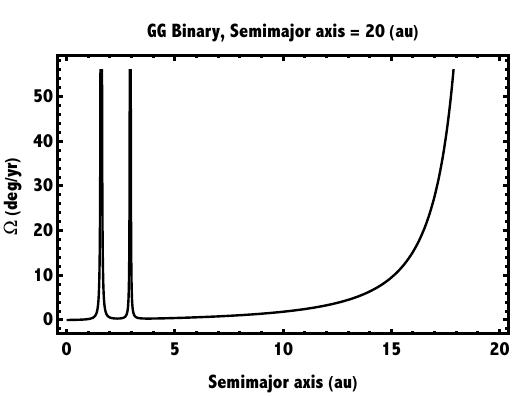}
\includegraphics[scale=0.9]{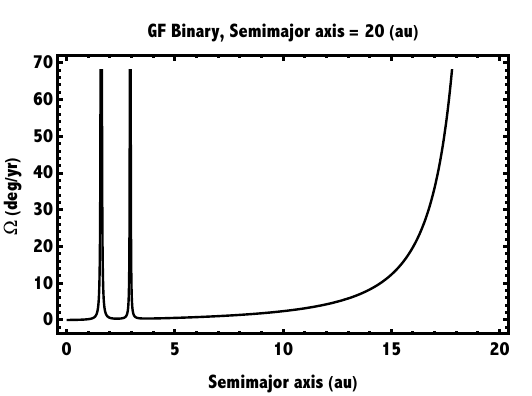}
\caption{Graphs of the variation of the frequency $\Omega_{\rm E}$ in terms of semimajor axis for a GM, GK, GG, and GF 
binary with a semimajor axis of 20 au. As expected, $\Omega_{\rm E}$ becomes singular at the locations of the two planets 
and the secondary star.}
\label{fig2}
\end{figure*}

\clearpage

\begin{figure*}[ht]
\hskip -15pt
{\includegraphics[scale=0.39]{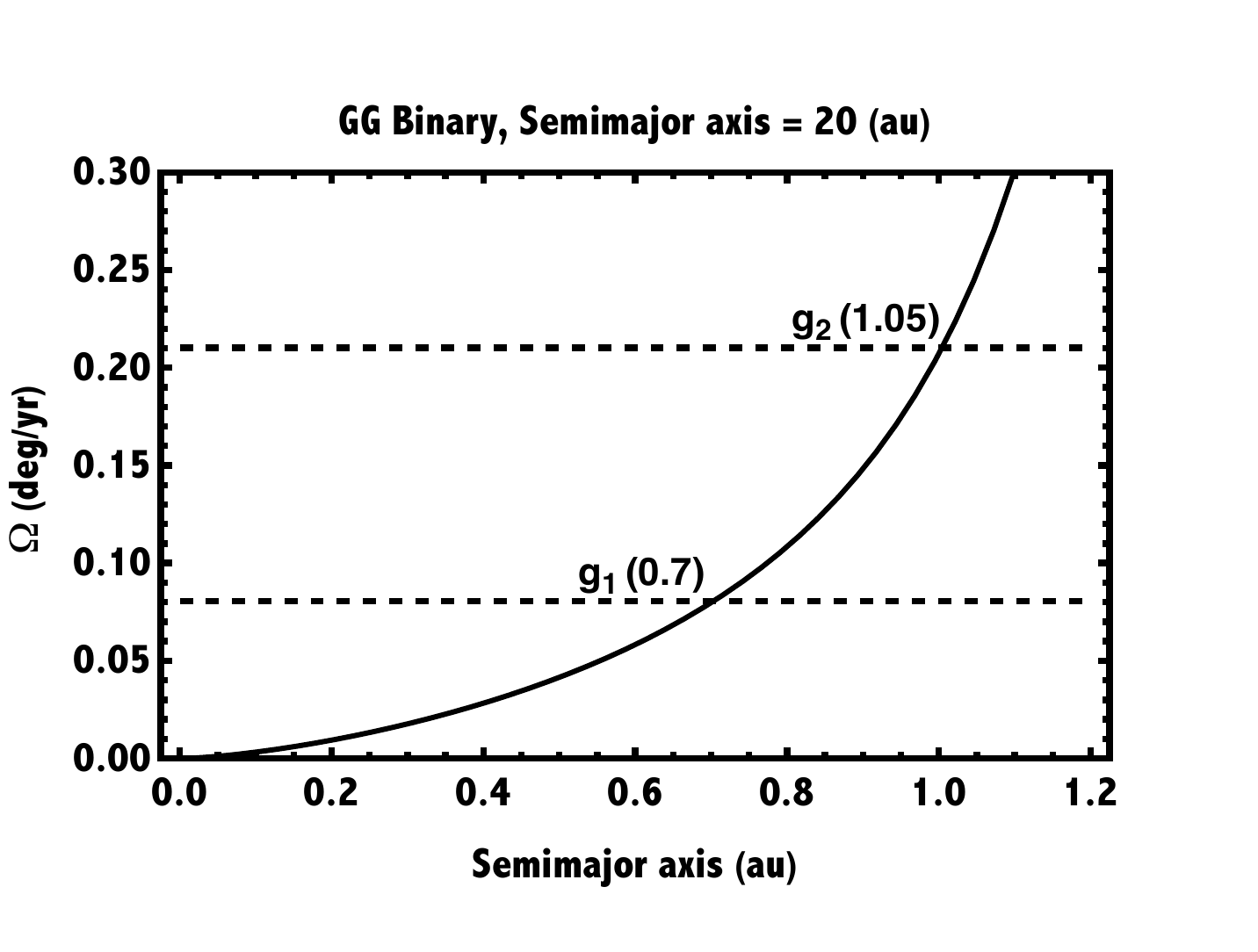}
\vskip -2.8in
\hskip 3.18in
\includegraphics[scale=0.39]{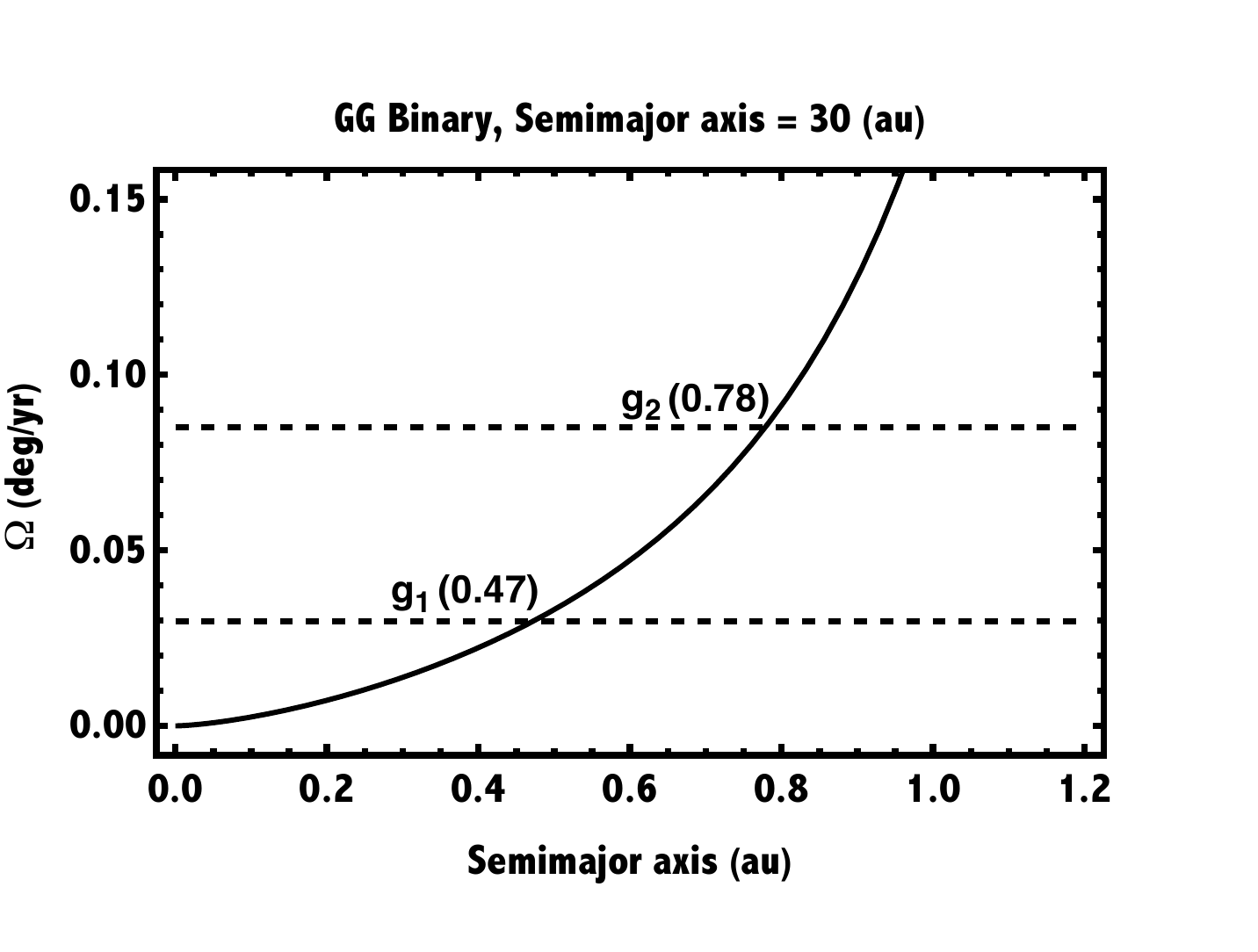}}
\vskip -15pt
\hskip -15pt
{\includegraphics[scale=0.39]{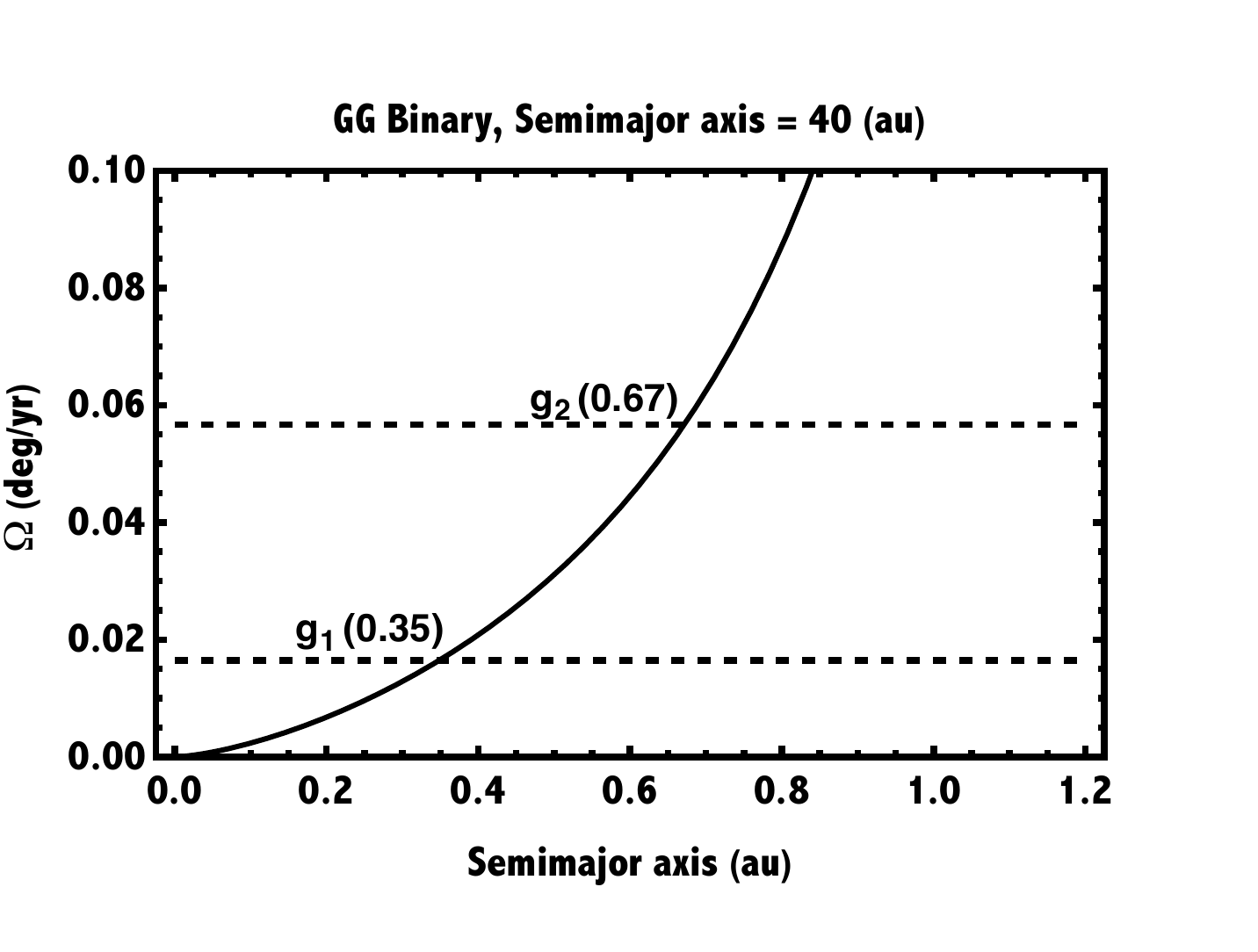}
\hskip -28pt
\includegraphics[scale=0.39]{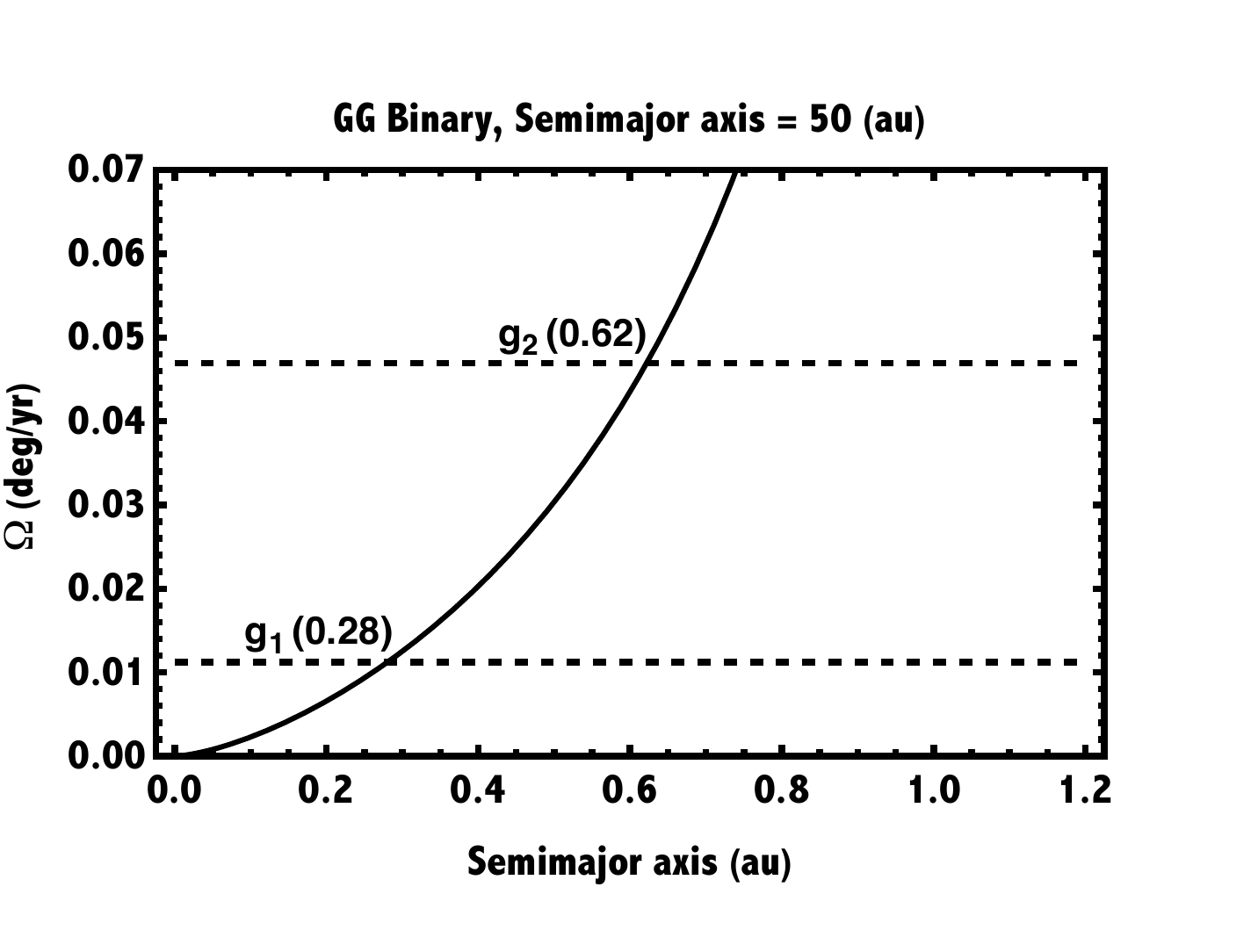}}
\vskip -15pt
\caption{Locations of the secular resonances of the inner and outer planets, $(g_1)$ and $(g_2)$, in a GG 
binary for different values of its semimajor axis.}
\label{fig3}
\end{figure*}

\clearpage

\begin{figure}
\vskip -10pt
\center{
\includegraphics[scale=0.43, trim = {0 0  0 2.0cm }, clip]{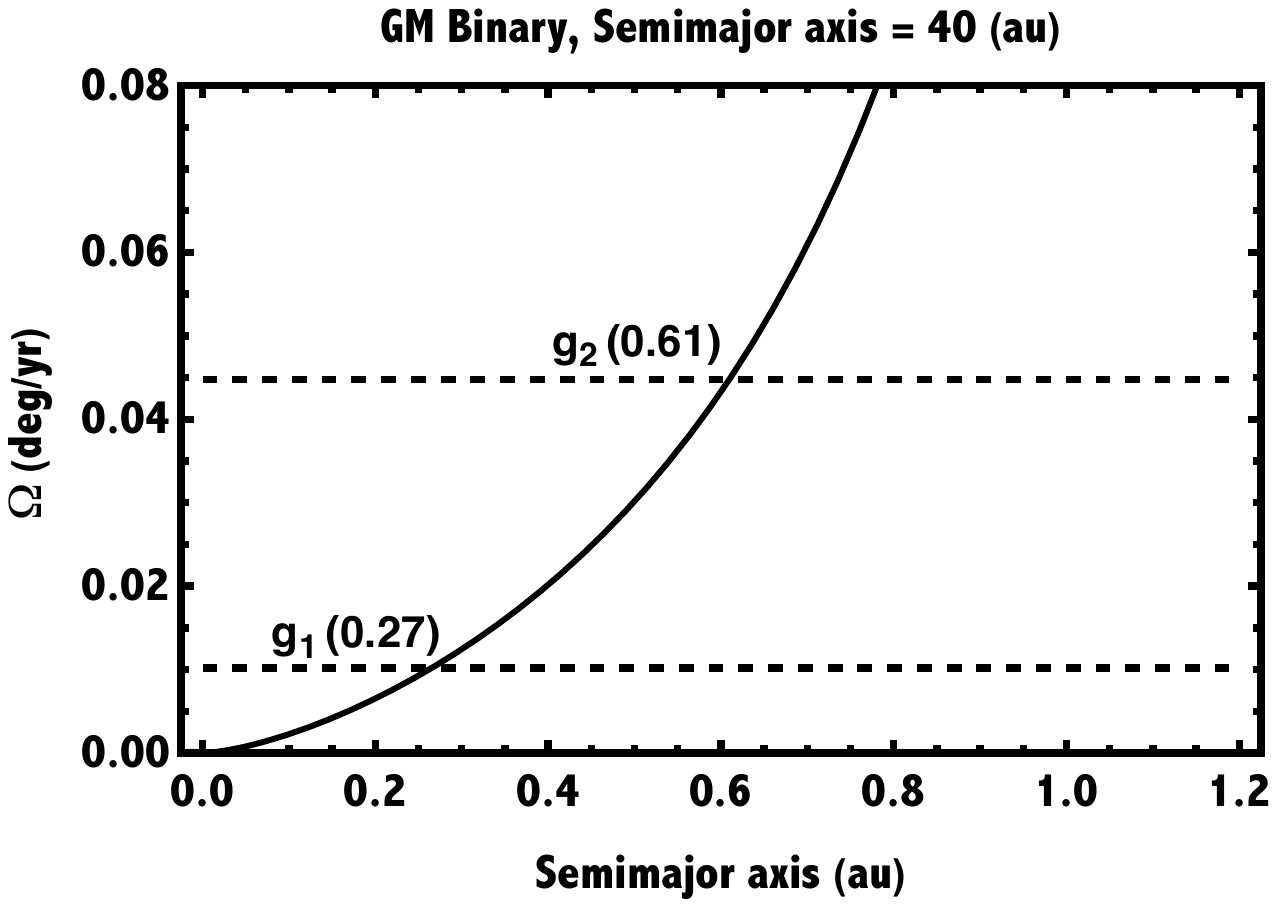}
\vskip -33pt
\includegraphics[scale=0.43, trim = {0 0 0 2.7cm}, clip]{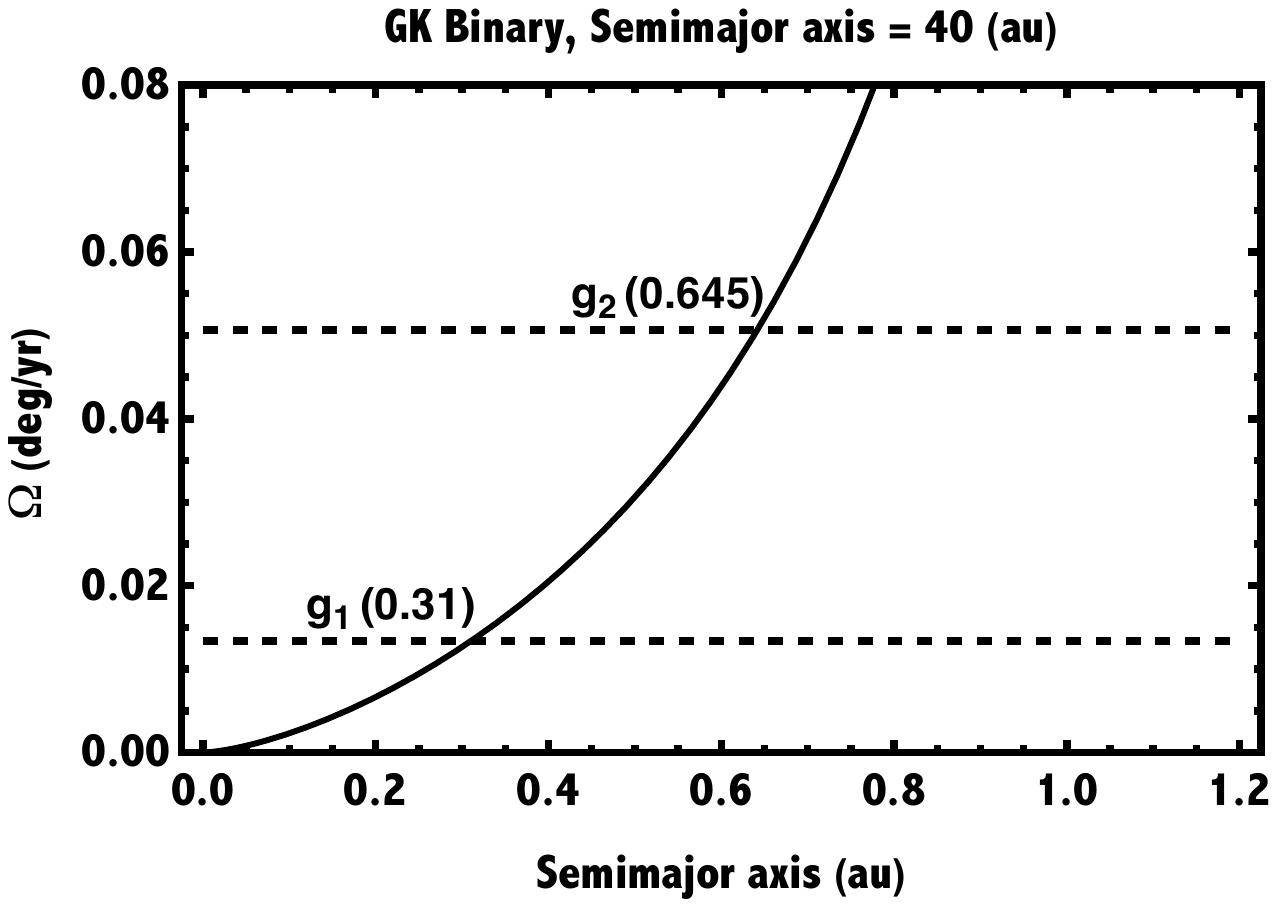}
\vskip -33pt
\includegraphics[scale=0.43, trim = {0 0 0 2.7cm}, clip]{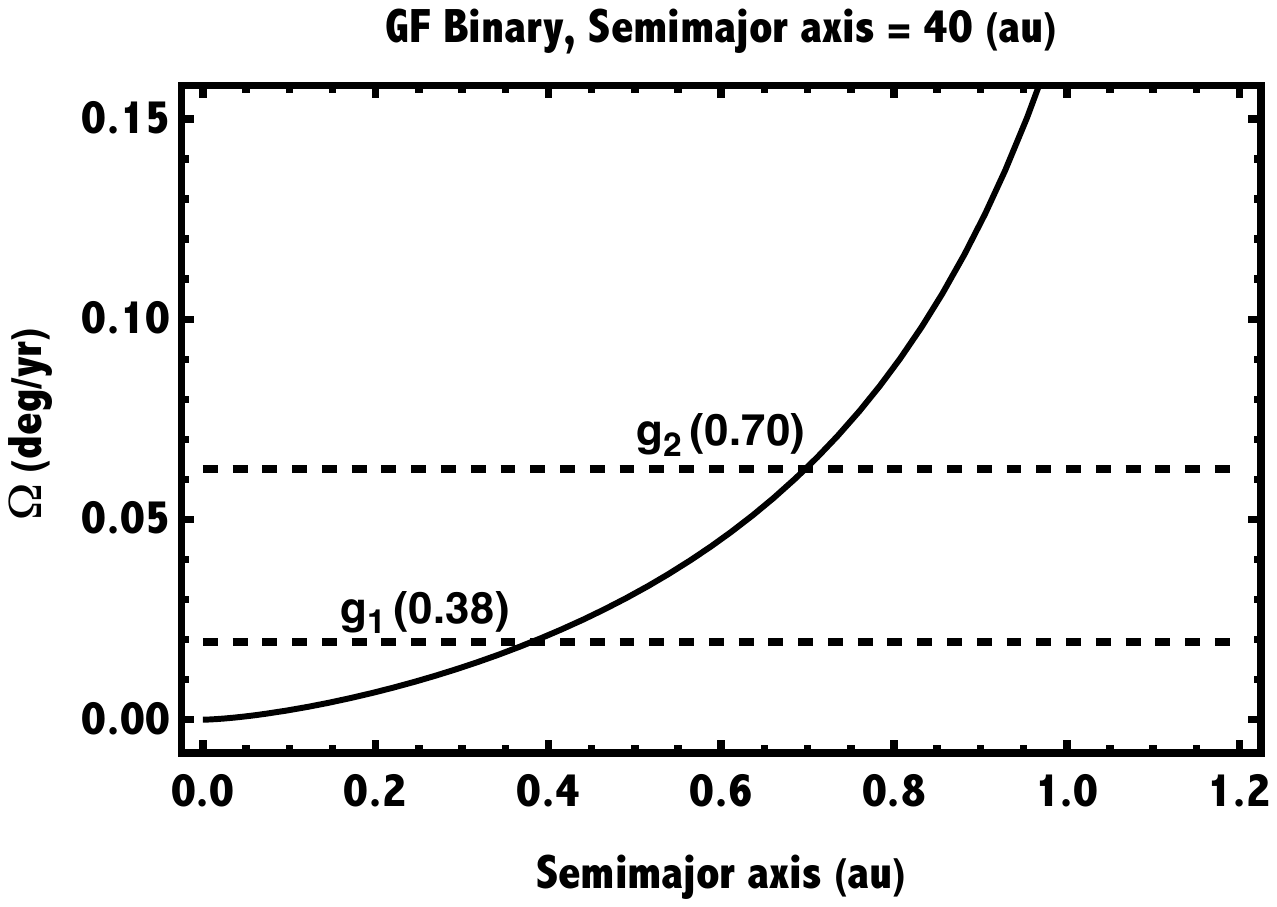}}
\vskip -45pt
\caption{Locations of the secular resonances of the inner and outer planets, $(g_1)$ and $(g_2)$, in a GM, GK,
and GF binary with a semimjor axis of 40 au.}
\label{fig4}
\end{figure}

\clearpage

\begin{figure*}[ht]
\vskip -20pt
\hskip -18pt
\includegraphics[scale=0.38]{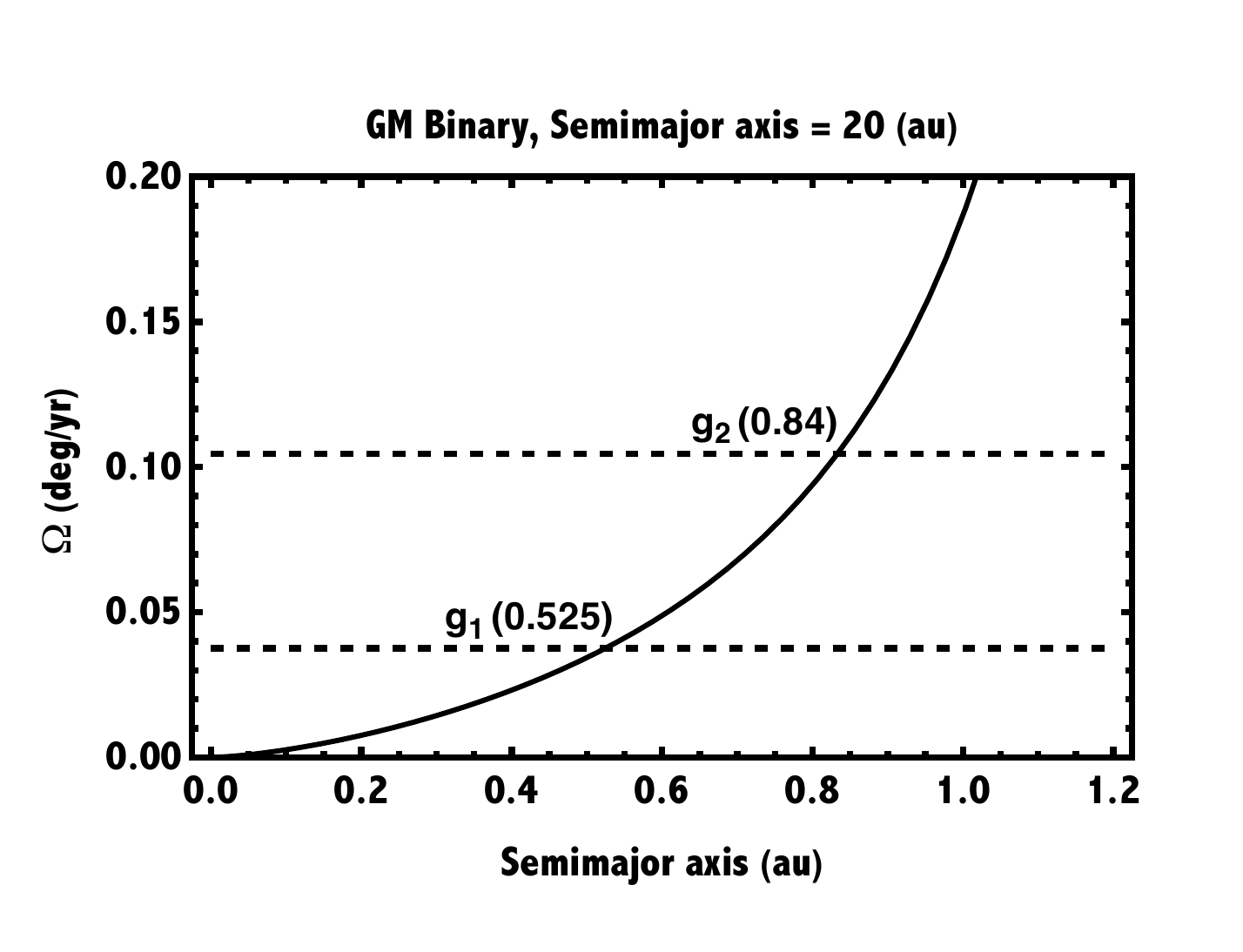}
\hskip -22pt
\includegraphics[scale=0.38]{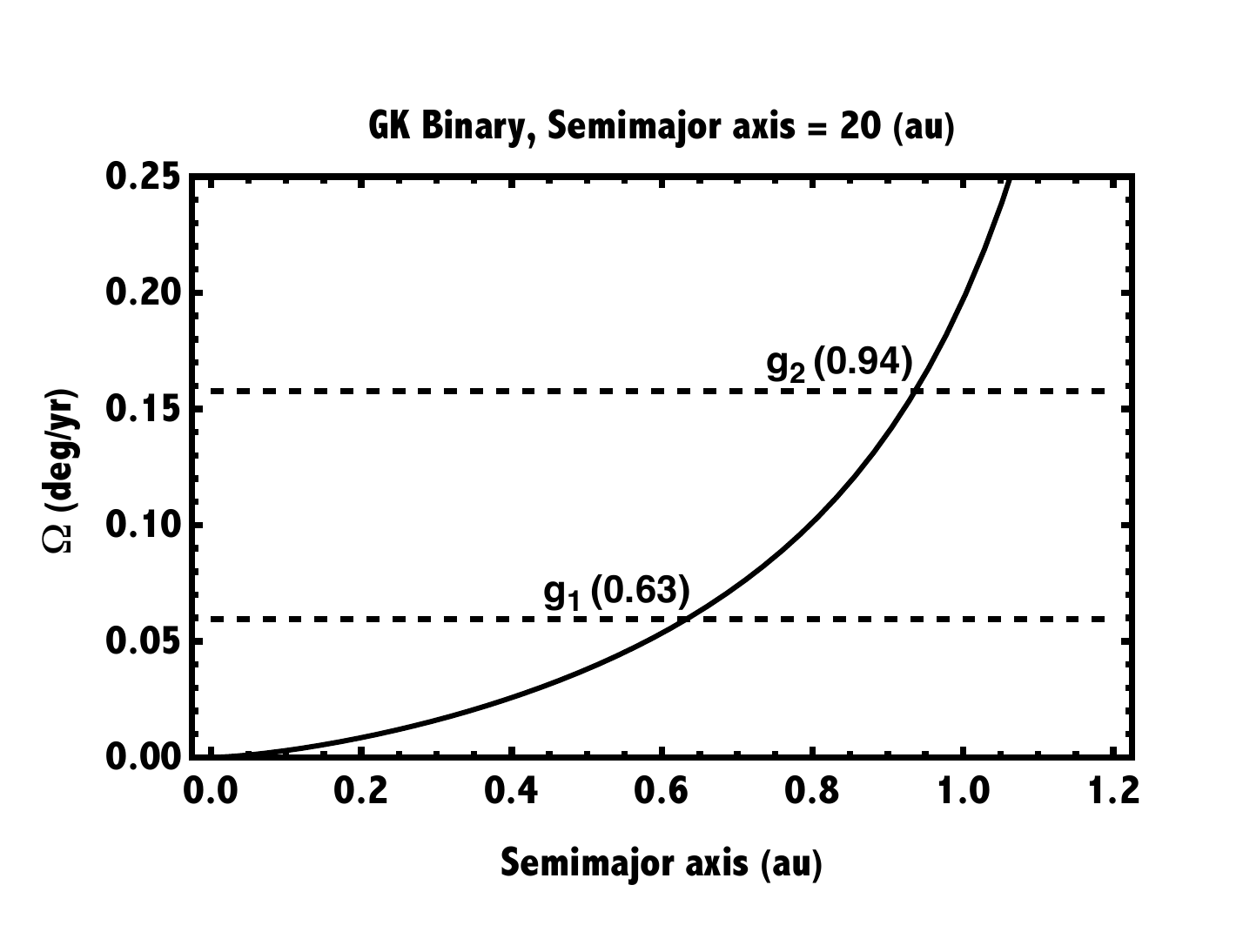}
\vskip  -43pt
\hskip -30pt
{\includegraphics[scale=0.68,  trim = {6.5cm 0 1.5cm 0}, clip]{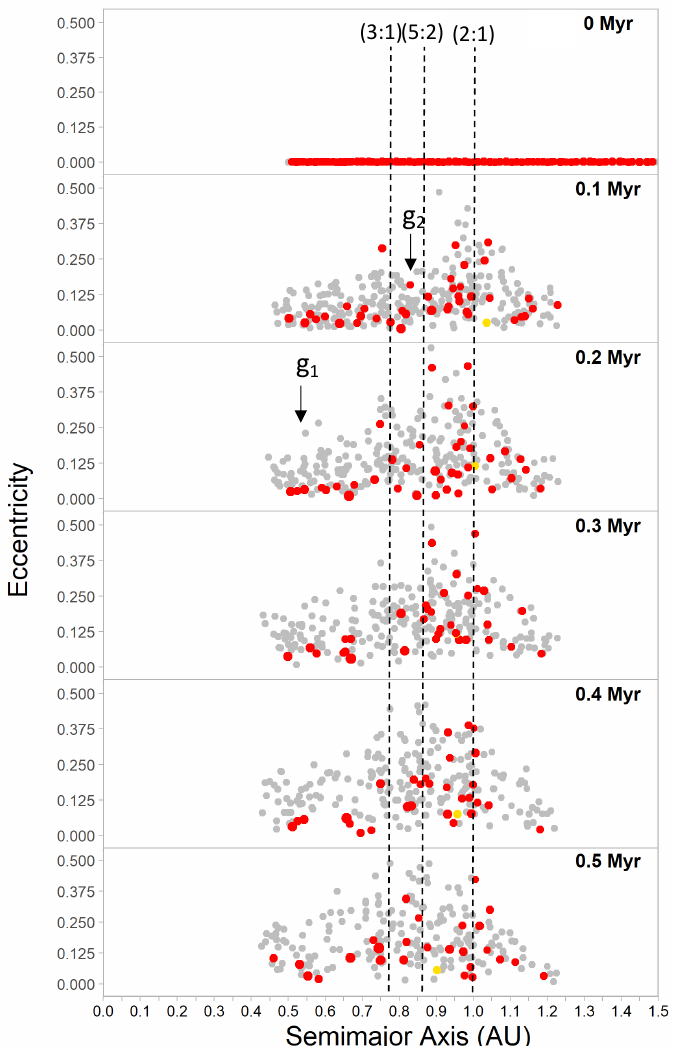}
\vskip -415pt
\hskip  227pt
\includegraphics[scale=0.68,  trim = {7.5cm 0 1.5cm 0}, clip]{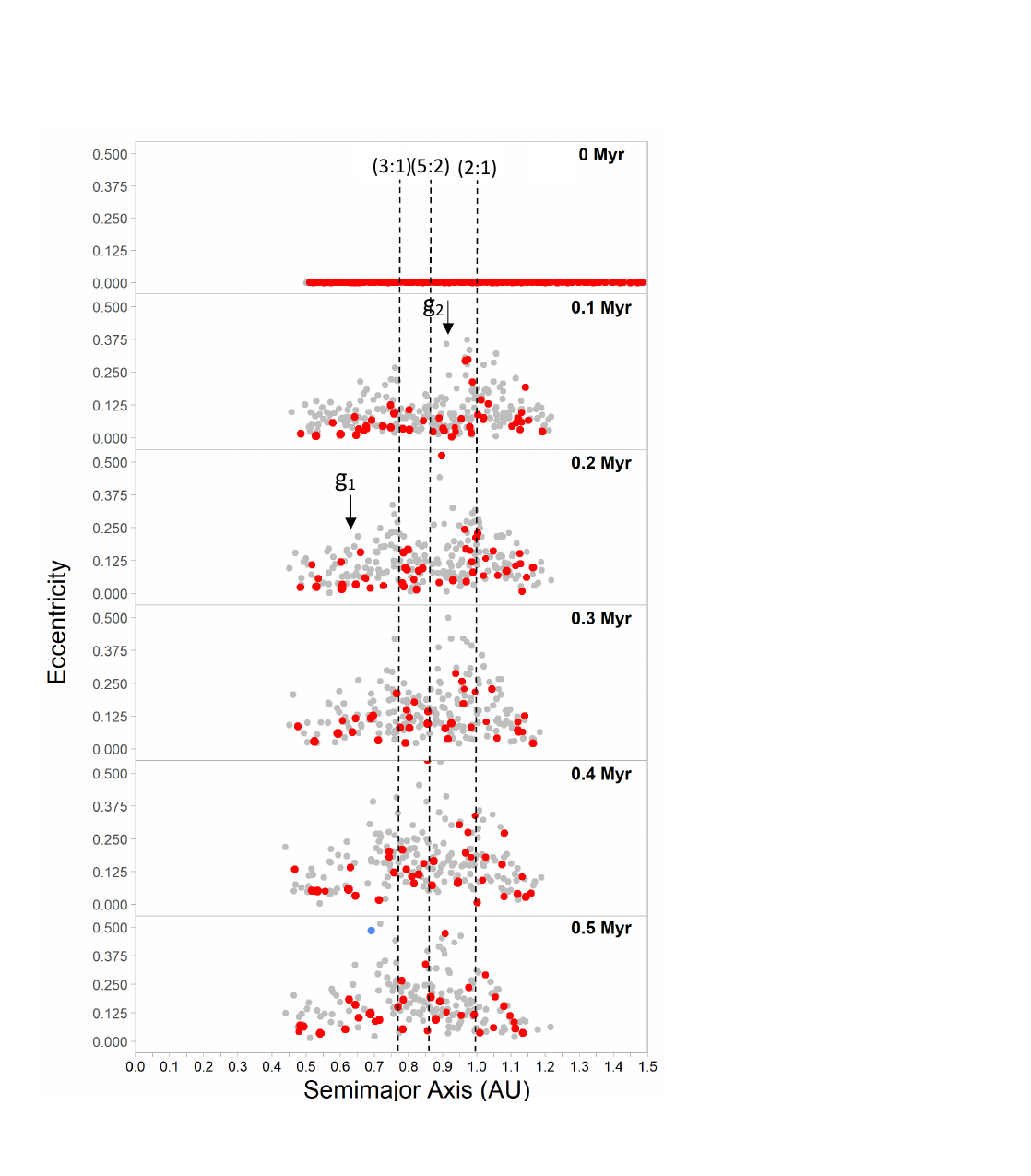}}
\vskip -30pt
\caption{Secular resonances in a GM and GK binary with a semimajor axis of 20 au. The top panels show the
locations of the secular resonances obtained from the general theory. The bottom panels show the secular resonances 
as they appear during the evolution of the protoplanetary disk. The red circles denote planetary embryos, gray circles are
planetesimals, and the dashed lines correspond to mean-motion resonances with the inner planet. Locations of secular resonances
are shown by arrows.}
\label{fig5}
\end{figure*}

\clearpage

\begin{figure*}[ht]
\vskip -20pt
\hskip -16pt
\includegraphics[scale=0.38]{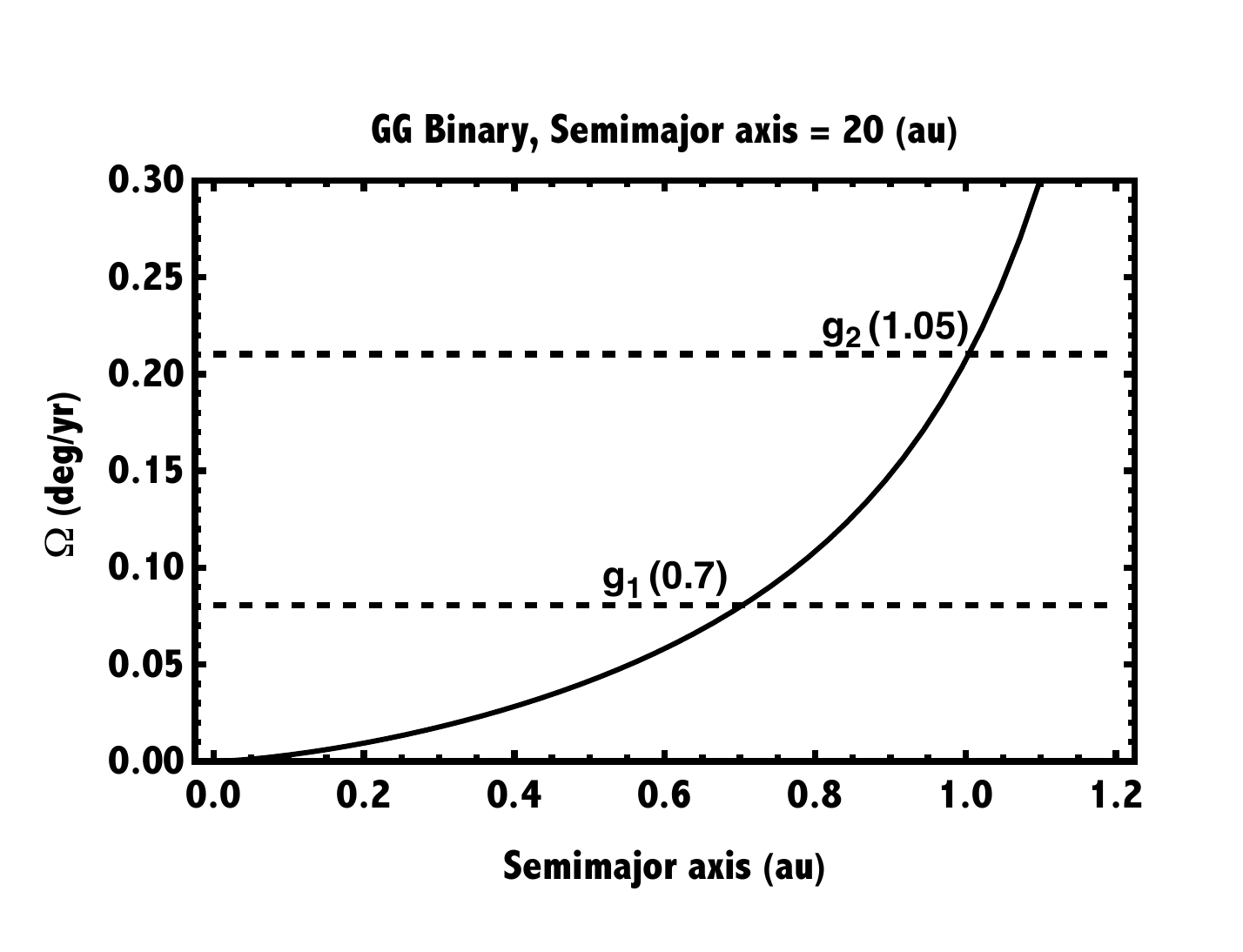}
\hskip -22pt
\includegraphics[scale=0.38]{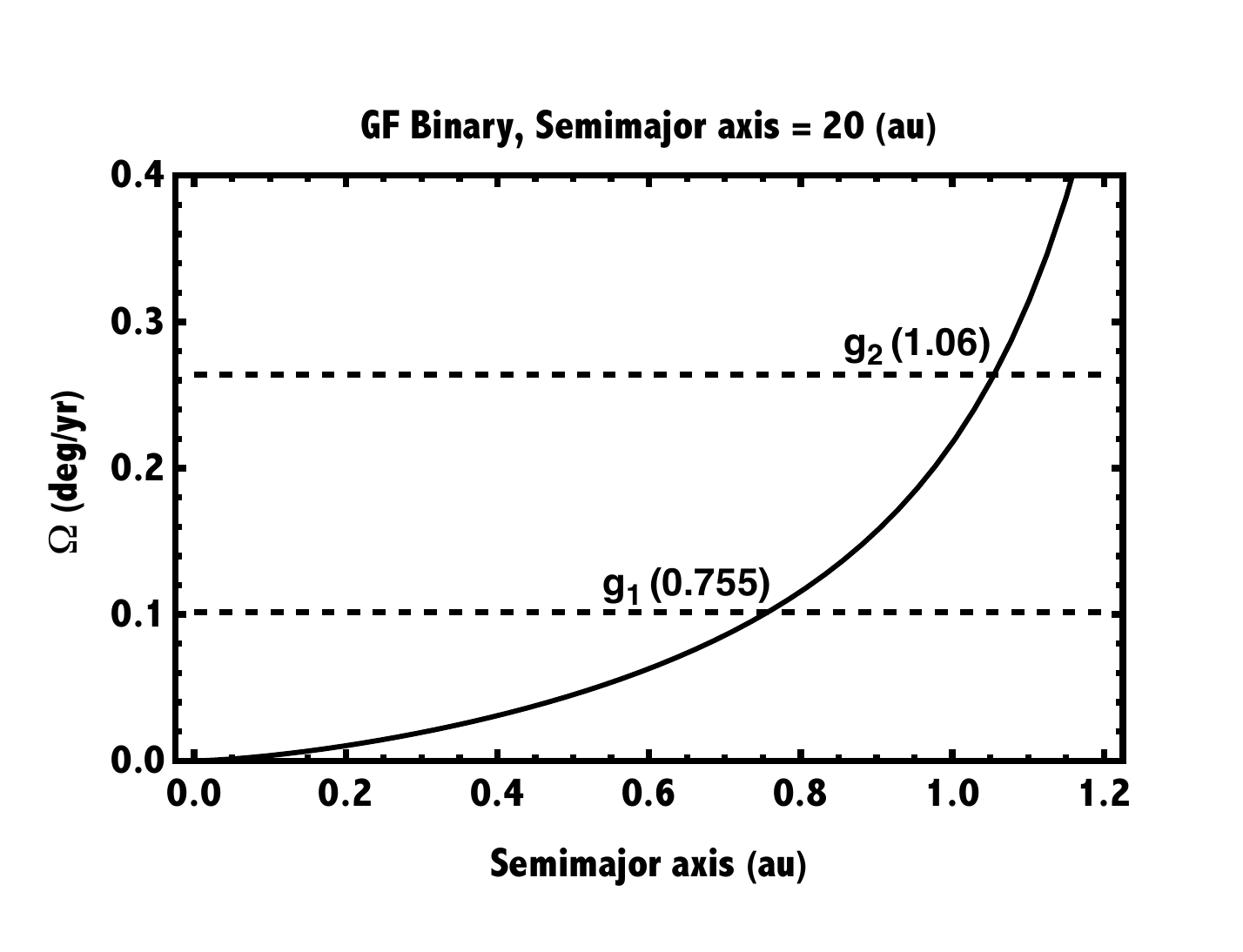}
\vskip  -43pt
\hskip -38pt
\includegraphics[scale=0.68,  trim = {6.5cm 0 1.5cm 0}, clip]{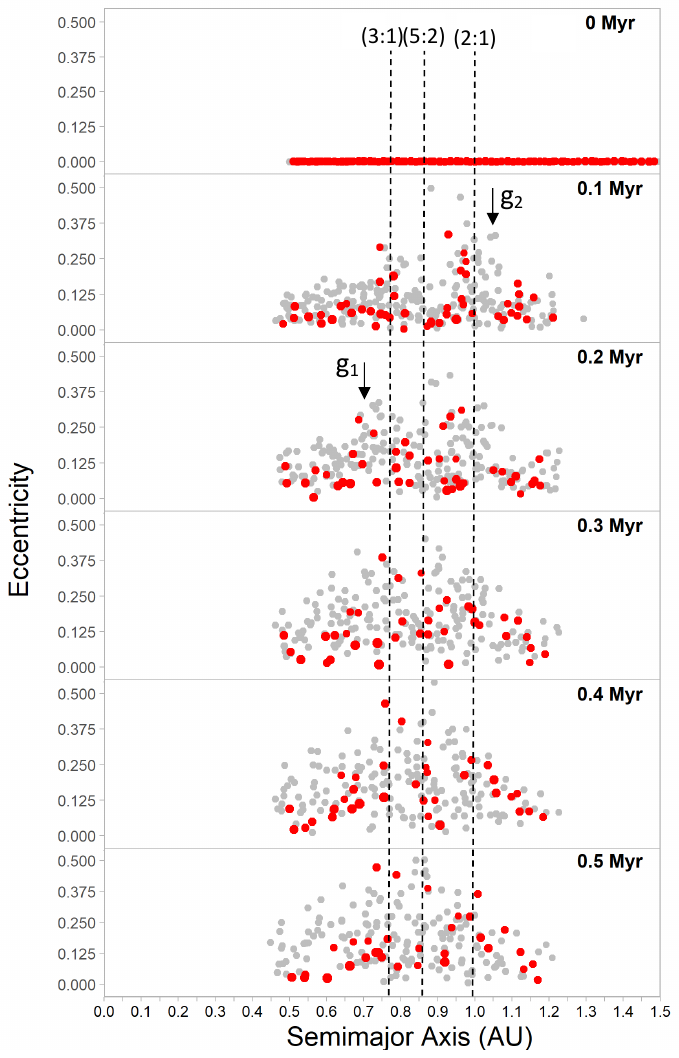}
\vskip -418pt
\hskip 235pt
\includegraphics[scale=0.68, trim = {7.5cm 0 1.5cm 0}, clip]{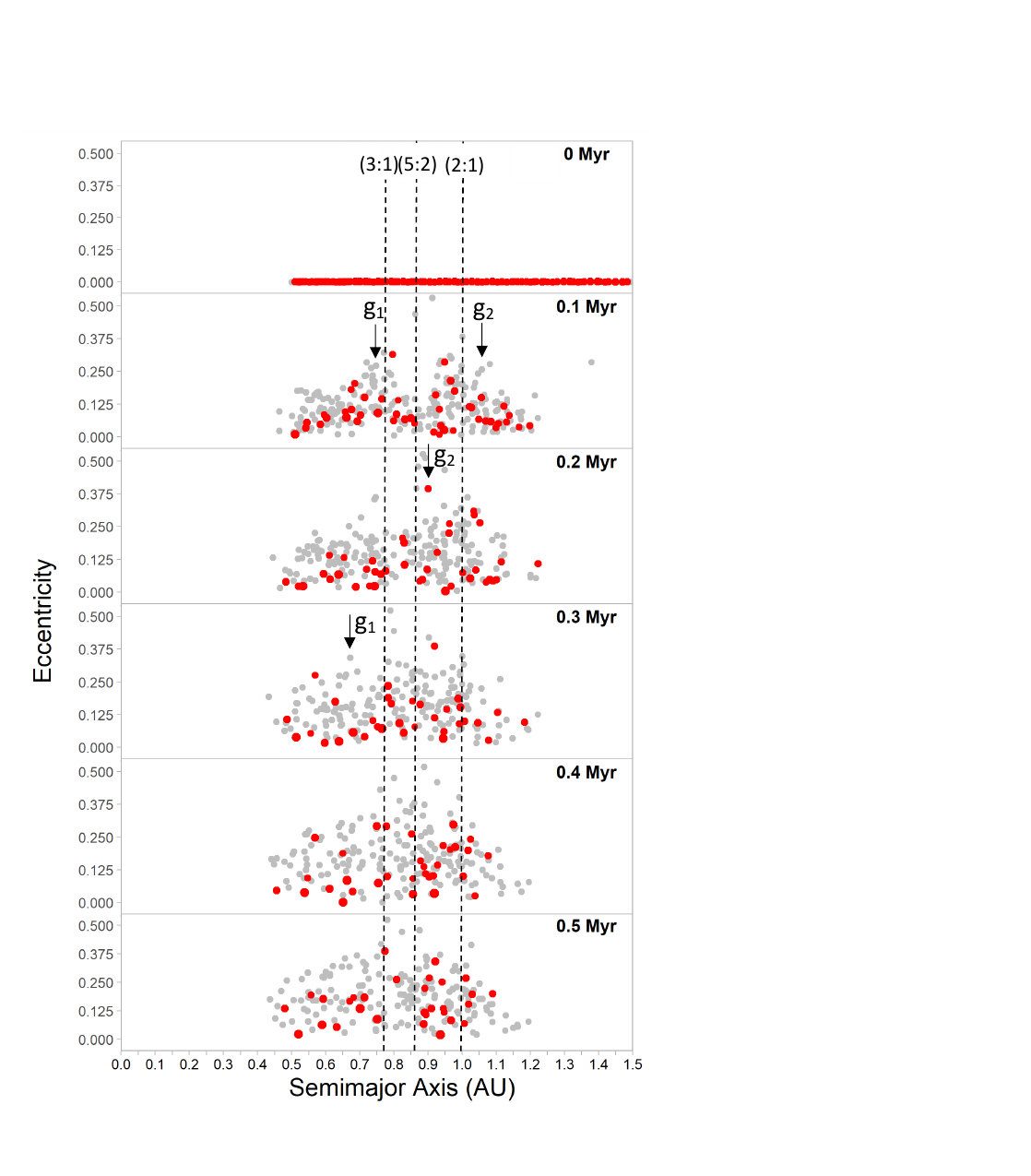}
\vskip -30pt
\caption{Secular resonances in a GG and GF binary with a semimajor axis of 20 au. The top panels show the
locations of the secular resonances obtained from the general theory. The bottom panels show the secular resonances 
as they appear during the evolution of the protoplanetary disk. The red circles denote planetary embryos, gray circles are
planetesimals, and the dashed lines correspond to mean-motion resonances with the inner planet. Locations of secular resonances
are shown by arrows.}
\label{fig6}
\end{figure*}

\clearpage

\begin{figure*}[ht]
\vskip  -10pt
\hskip -27pt
\includegraphics[scale=0.7, trim = {7cm 0 1.5cm 0}, clip]{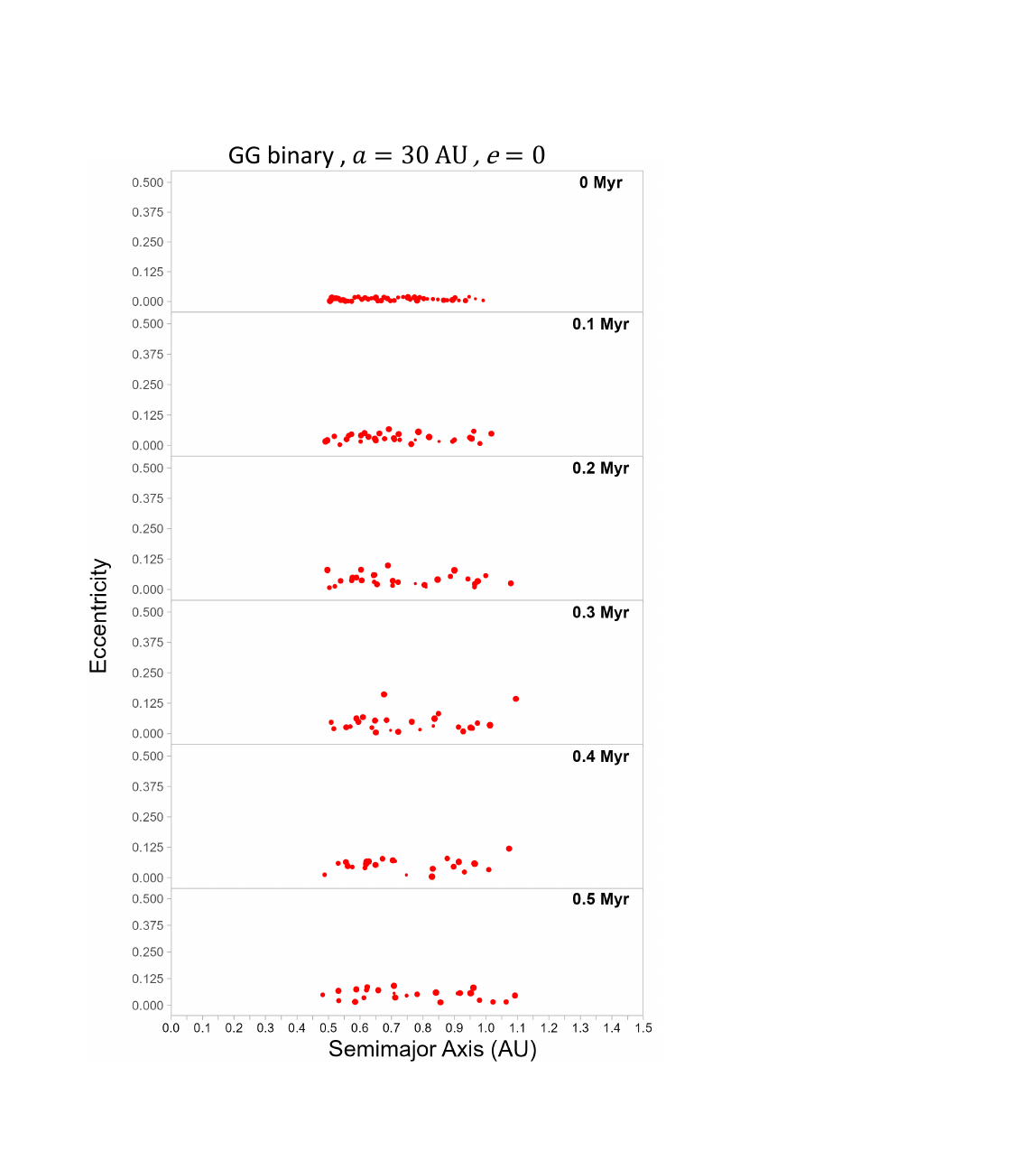}
\vskip -430pt
\hskip 215pt
\includegraphics[scale=0.7, trim = {7cm 0 1.5cm 0}, clip]{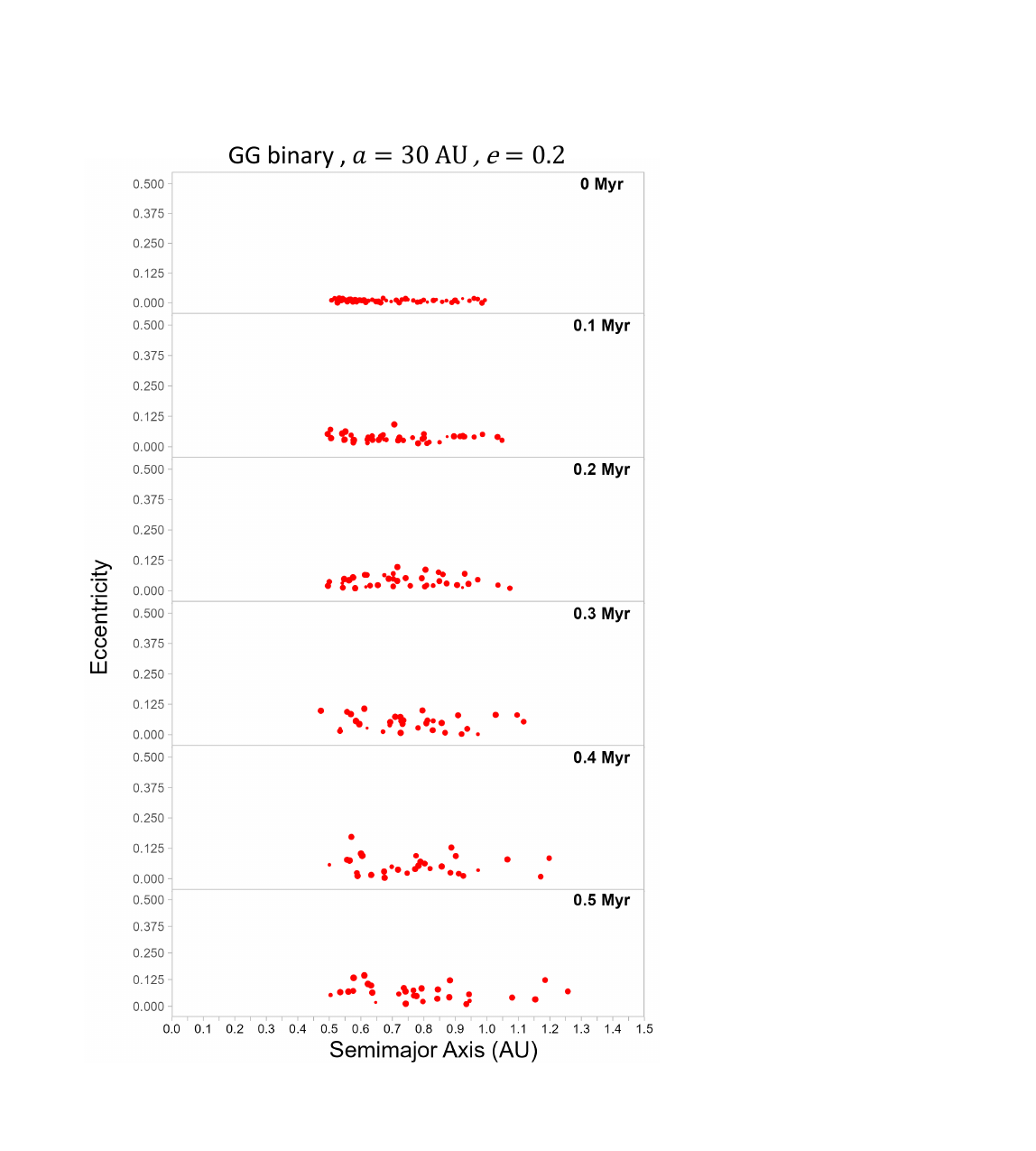}
\vskip -40pt
\caption{Graph of the evolution of the primary's protoplanetry disk in a binary without giant planets. As shown here,
because there are no giant planets, no secular resonances appear in the disk.}
\label{fig7}
\end{figure*}

\clearpage

\begin{figure}
\vskip -40pt
\center{
\includegraphics[scale=0.45]{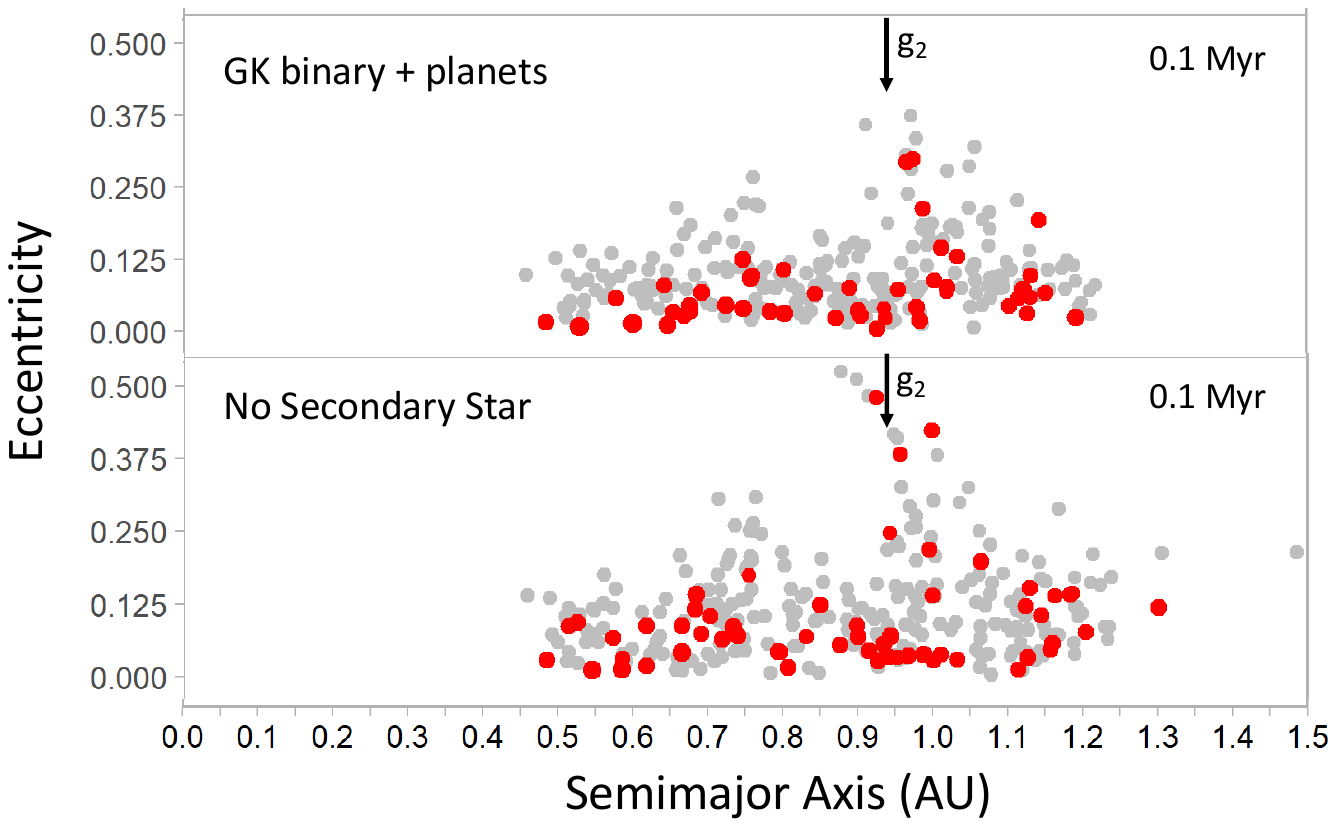}
\vskip -70pt
\includegraphics[scale=0.45]{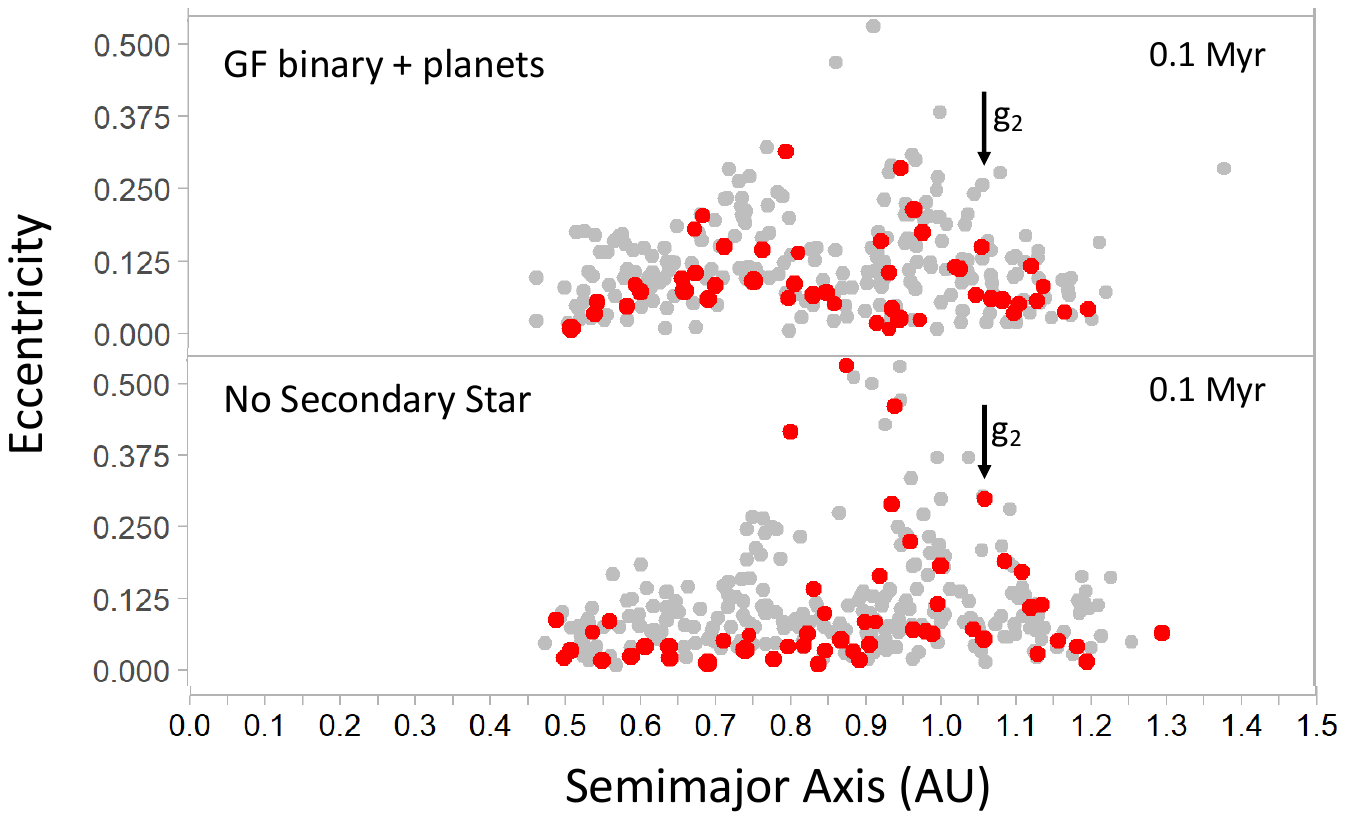}}
\vskip -60pt
\caption{Locations of secular resonances in a 20 au GK (top) and GF (bottom) binary with and without the secodnary star.
As shown here, the orbital excitation of the planetary embryos (the red circles) are larger in systems without
the secondary star.}
\label{fig8}
\end{figure}

\clearpage

\begin{figure*}[ht]
\hskip -40pt
{\includegraphics[scale=0.4]{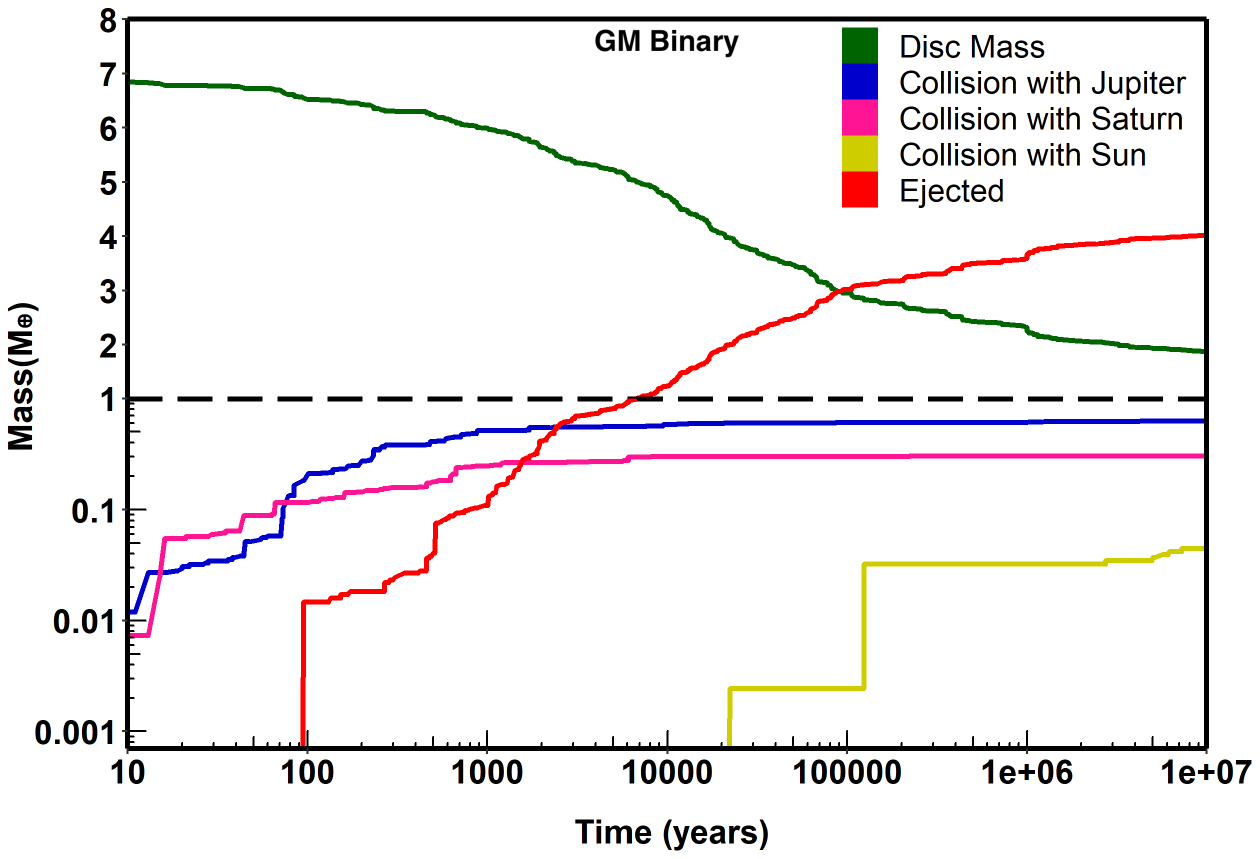}
\vskip -3.3in
\hskip 3.1in
\includegraphics[scale=0.4, trim = {1cm 1.3cm 0 0.5cm}, clip]{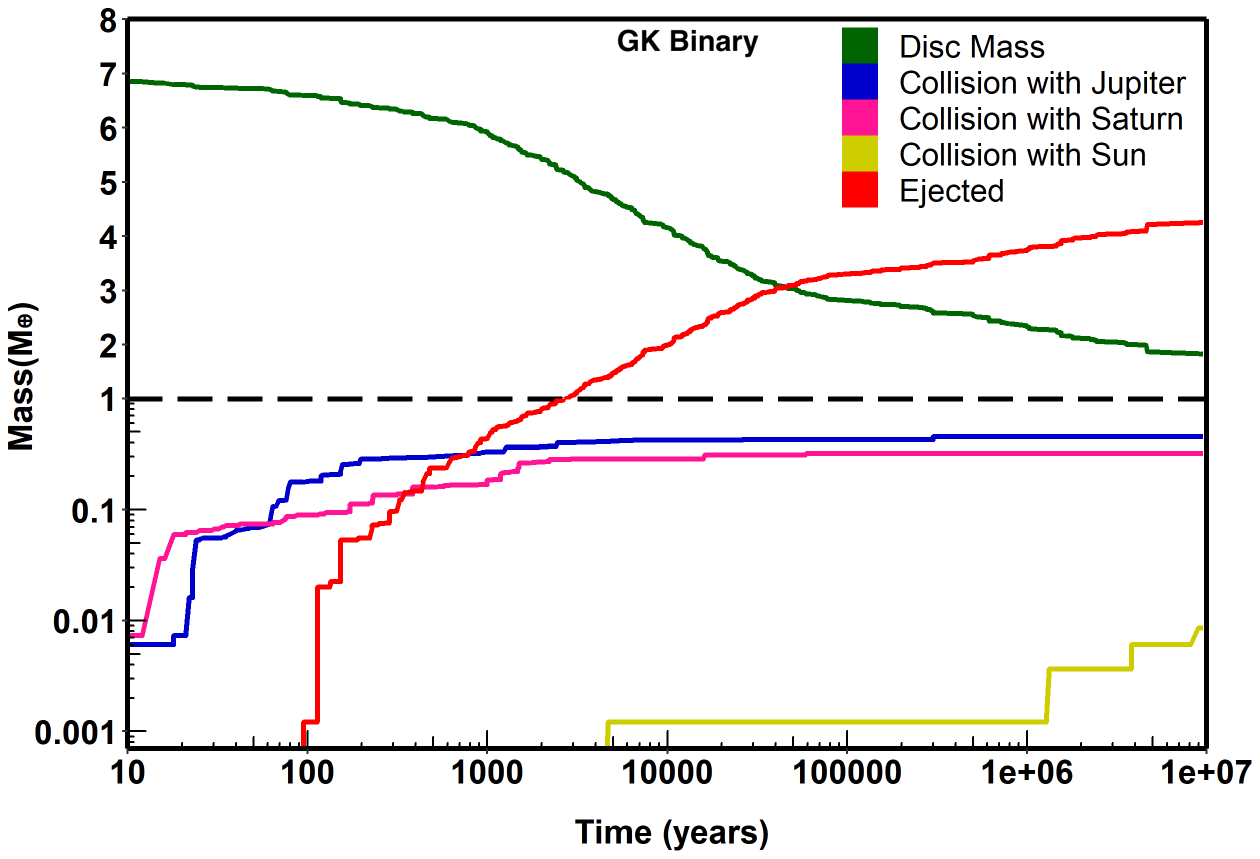}}
\vskip -45pt
\hskip -40pt
{\includegraphics[scale=0.4]{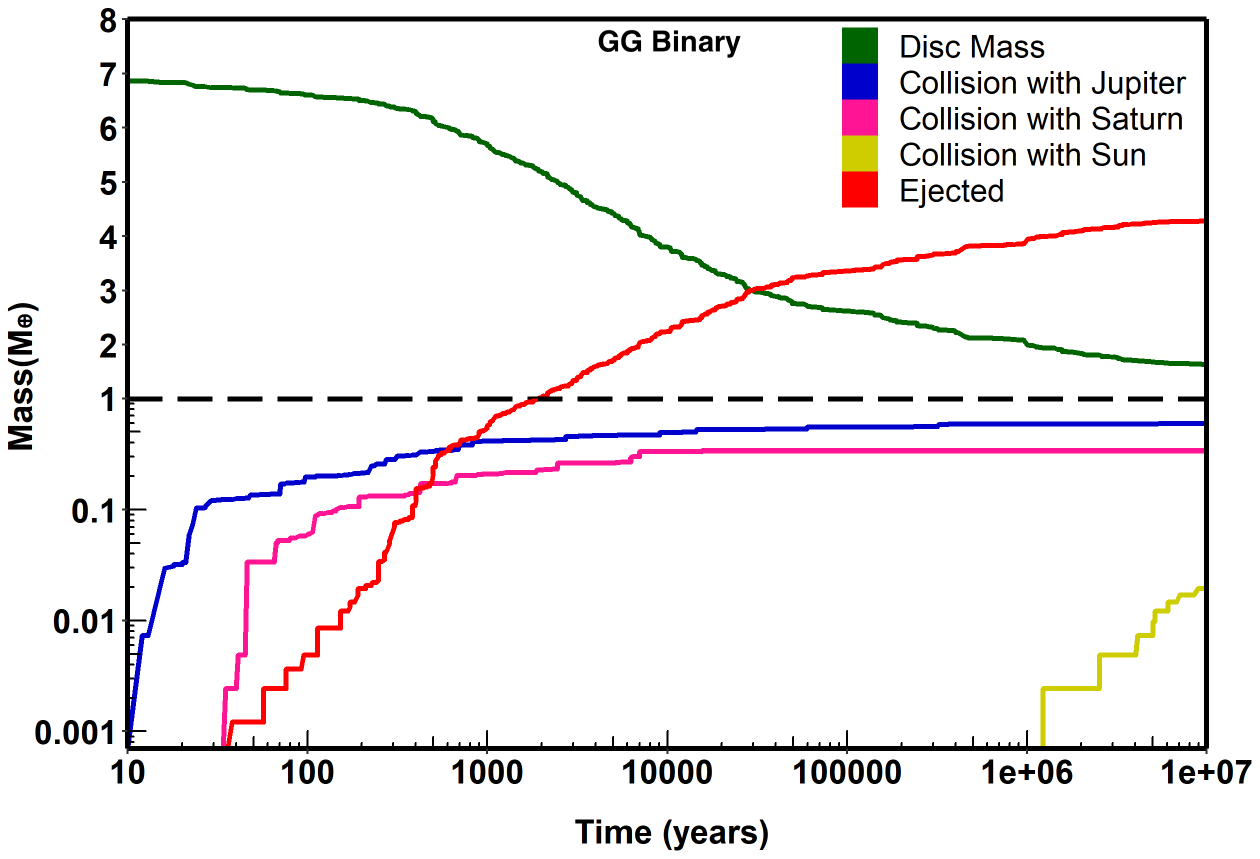}
\hskip -70pt
\includegraphics[scale=0.4]{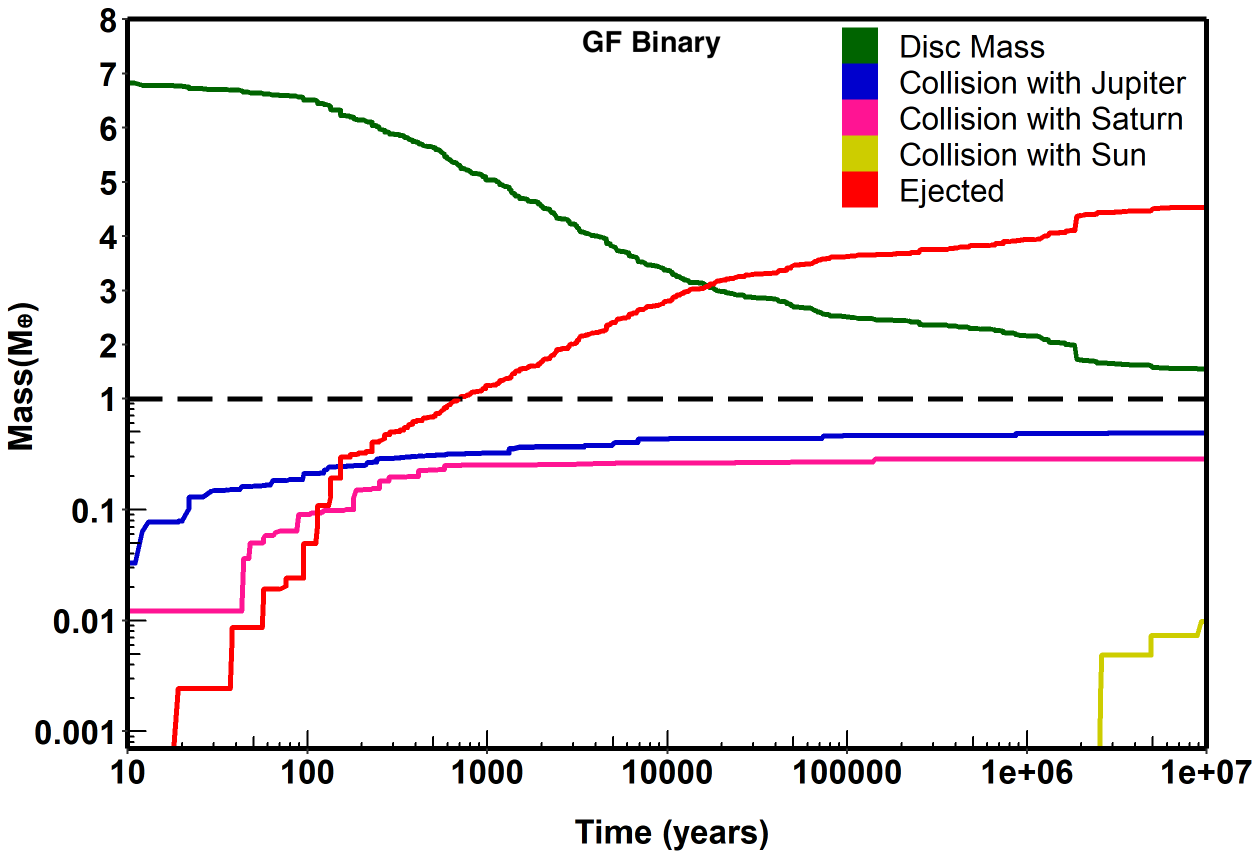}}
\vskip -40pt
\caption{The evolution of the mass of the protoplanetary disk in the binaries of Figures 5 and 6. As shown here, 
the majority of the mass loss is through ejection which begins earlier as the mass of the secondary increases. 
Note that the rate of collision with the primary star (denoted by the word ``Sun'') decreases and moves to
later times for binaries with more massive secondary stars.}
\label{fig9}
\end{figure*}

\clearpage

\begin{deluxetable}{c|c|c|c|c}
\tabletypesize{\scriptsize}
\tablecaption{Secondary: M star (0.4 Solar mass)}
\label{table1}
\tablewidth{0pt}
\tablehead{\colhead{$A_{ij}$} & \multicolumn{4}{c}{Semimajor Axis} \\
\hline
\colhead {} & \colhead{20 (au)} & \colhead {30 (au)} &  \colhead{40 (au)} & \colhead {50 (au)}}
 \startdata
$A_{11}$ & 0.0402  & 0.0207  &  0.0159 & 0.0143 \\
$A_{12}$ & -0.0082  & -0.0082  & -0.0082 & -0.0082 \\
$A_{13}$ & -0.0028  & -0.0005  & -0.0002  & 0.0001 \\
$A_{21}$ & -0.0202  & -0.0202   & -0.0202   & -0.0202 \\
$A_{22}$ & 0.1020 & 0.0510 & 0.039 & -0.0348 \\
$A_{23}$ & -0.0130 & -0.0025 & -0.0008 & -0.0003 \\
$A_{31}$ & $-1.66 \times {10^{-6}}$ & $-2.66 \times {10^{-7}}$ & $-7.30 \times {10^{-8}}$ & $-2.60 \times {10^{-8}}$ \\
$A_{32}$ & $-3.20 \times {10^{-6}}$ & $-5.10 \times {10^{-7}}$ & $-1.37 \times {10^{-7}}$ & $-5.10 \times {10^{-8}}$ \\
$A_{33}$ & $-3.38 \times {10^{-5}}$ & $8.00 \times {10^{-6}}$ & $3.00 \times {10^{-6}}$ & $1.35 \times {10^{-6}}$ \\
\hline
$g_{1}$ & 0.104573 & 0.055725 & 0.044749 & 0.040995 \\
$g_{2}$ & 0.037604 & 0.015945 & 0.010191 & 0.008065 \\
$g_{3}$ &  $3.3163 \times {10^{-5}}$ & $7.9503 \times {10^{-6}}$ & $2.9930 \times {10^{-6}}$ & $1.3487 \times {10^{-6}}$ \\

\enddata
\end{deluxetable}

\begin{deluxetable}{c|c|c|c|c}
\tabletypesize{\scriptsize}
\tablecaption{Secondary: K star (0.7 Solar mass)}
\label{table2}
\tablewidth{0pt}
\tablehead{\colhead{$A_{ij}$} & \multicolumn{4}{c}{Semimajor Axis} \\
\hline
\colhead {} & \colhead{20 (au)} & \colhead {30 (au)} &  \colhead{40 (au)} & \colhead {50 (au)}}
 \startdata
$A_{11}$ & 0.0612  & 0.0268  &  0.0185    & 0.0156 \\
$A_{12}$ & -0.0082  & -0.0082  & -0.0082  & -0.0082 \\
$A_{13}$ & -0.0048  & -0.0009  & -0.0003  & -0.0001 \\
$A_{21}$ & -0.0202  & -0.0202  & -0.0202  & -0.0202 \\
$A_{22}$ & 0.156    & 0.0644   & 0.0455   & 0.0380 \\
$A_{23}$ & -0.023   & -0.0044 & -0.0014 & -0.0006 \\
$A_{31}$ & $-1.50 \times {10^{-6}}$ & $-2.40 \times {10^{-7}}$ & $-6.60 \times {10^{-8}}$ & $-2.40 \times {10^{-8}}$ \\
$A_{32}$ & $-2.88 \times {10^{-6}}$ & $-4.60 \times {10^{-7}}$ & $-1.20 \times {10^{-7}}$ & $-4.60 \times {10^{-8}}$ \\
$A_{33}$ & $-3.10 \times {10^{-5}}$ & $7.33 \times {10^{-6}}$  & $2.70 \times {10^{-6}}$  & $1.23 \times {10^{-6}}$ \\
\hline
$g_{1}$ & 0.157716  & 0.0702153 & 0.050655 & 0.043857 \\
$g_{2}$ & 0.0594416 & 0.0229848 & 0.013365 & 0.009720 \\
$g_{3}$ &  $3.0388 \times {10^{-5}}$ & $7.2763 \times {10^{-6}}$ & $2.6920 \times {10^{-6}}$ & $1.2283 \times {10^{-6}}$ \\

\enddata
\end{deluxetable}

\clearpage

\begin{deluxetable}{c|c|c|c|c}
\tabletypesize{\scriptsize}
\tablecaption{Secondary: G star (1.0 Solar mass)}
\label{table3}
\tablewidth{0pt}
\tablehead{\colhead{$A_{ij}$} & \multicolumn{4}{c}{Semimajor Axis} \\
\hline
\colhead {} & \colhead{20 (au)} & \colhead {30 (au)} &  \colhead{40 (au)} & \colhead {50 (au)}}
 \startdata
$A_{11}$ & 0.0819  & 0.0330  &  0.0211    & 0.0167 \\
$A_{12}$ & -0.0082  & -0.0082  & -0.0082  & -0.0082 \\
$A_{13}$ & -0.0007  & -0.0013  & -0.0004  & -0.0002 \\
$A_{21}$ & -0.0202  & -0.0202  & -0.0202  & -0.0202 \\
$A_{22}$ & 0.2090    & 0.0818   & 0.0520   & 0.0414 \\
$A_{23}$ & -0.0326   & -0.0063 & -0.0020 & -0.0008 \\
$A_{31}$ & $-1.38 \times {10^{-6}}$ & $-2.20 \times {10^{-7}}$ & $-6.10 \times {10^{-8}}$ & $-2.20 \times {10^{-8}}$ \\
$A_{32}$ & $-2.66 \times {10^{-6}}$ & $-4.20 \times {10^{-7}}$ & $-1.10 \times {10^{-7}}$ & $-4.20 \times {10^{-8}}$ \\
$A_{33}$ & $2.83 \times {10^{-5}}$ & $6.76 \times {10^{-6}}$  & $2.50 \times {10^{-6}}$  & $1.10 \times {10^{-6}}$ \\
\hline
$g_{1}$ & 0.2102910  & 0.084986 & 0.056658 & 0.046921 \\
$g_{2}$ & 0.080610   & 0.029814 & 0.016442 & 0.011219 \\
$g_{3}$ &  $2.7718 \times {10^{-5}}$ & $6.71 \times {10^{-6}}$ & $2.49 \times {10^{-6}}$ & $2.00 \times {10^{-6}}$ \\

\enddata
\end{deluxetable}

\begin{deluxetable}{c|c|c|c|c}
\tabletypesize{\scriptsize}
\tablecaption{Secondary: F star (1.3 Solar-Mass)}
\label{table4}
\tablewidth{0pt}
\tablehead{\colhead{$A_{ij}$} & \multicolumn{4}{c}{Semimajor Axis} \\
\hline
\colhead {} & \colhead{20 (au)} & \colhead {30 (au)} &  \colhead{40 (au)} & \colhead {50 (au)}}
\startdata
$A_{11}$ & 0.1072  & 0.0390  &  0.0237    & 0.0182 \\
$A_{12}$ & -0.0082  & -0.0082  & -0.0082  & -0.0082 \\
$A_{13}$ & -0.0090  & -0.0174  & -0.0006  & -0.0002 \\
$A_{21}$ & -0.0202  & -0.0202  & -0.0202  & -0.0202 \\
$A_{22}$ & 0.2630    & 0.0973   & 0.0584   & 0.0447 \\
$A_{23}$ & -0.0424   & -0.0082 & -0.0026 & -0.0010 \\
$A_{31}$ & $-1.30 \times {10^{-6}}$ & $-2.00 \times {10^{-7}}$ & $-5.70 \times {10^{-8}}$ & $-2.00 \times {10^{-8}}$ \\
$A_{32}$ & $-2.50 \times {10^{-6}}$ & $-4.00 \times {10^{-7}}$ & $-1.00 \times {10^{-7}}$ & $-4.00 \times {10^{-8}}$ \\
$A_{33}$ & $2.64 \times {10^{-5}}$ & $6.30 \times {10^{-6}}$  & $2.30 \times {10^{-6}}$  & $1.00 \times {10^{-6}}$ \\
\hline
$g_{1}$ & 0.264027 & 0.100015  & 0.062649 & 0.049924 \\
$g_{2}$ & 0.101693   & 0.036285  & 0.019421 & 0.012991 \\
$g_{3}$ &  $2.58 \times {10^{-5}}$ & $6.13 \times {10^{-6}}$ & $2.29 \times {10^{-6}}$ & $1.00 \times {10^{-6}}$ \\

\enddata
\end{deluxetable}

\end{document}